\documentclass[twocolumn]{aastex61}

\graphicspath{{figs/}}
\usepackage{xspace}
\usepackage{booktabs}
\usepackage{bm}
\usepackage{romannum}
\usepackage{array}
\usepackage[counterclockwise, figuresright]{rotating}
\usepackage[utf8]{inputenc}
\def\nsam{\mbox{30}\xspace}

\def\slope{\mbox{0.58}\xspace}
\def\slopedev{\mbox{0.11}\xspace}
\def\intercept{\mbox{6.69}\xspace}
\def\interceptdev{\mbox{0.10}\xspace}

\def\gs{\mathrel{\raise0.35ex\hbox{$\scriptstyle >$}\kern-0.6em \lower0.40ex\hbox{{$\scriptstyle \sim$}}}}
\def\ls{\mathrel{\raise0.35ex\hbox{$\scriptstyle <$}\kern-0.6em \lower0.40ex\hbox{{$\scriptstyle \sim$}}}}

\def\oi{\mbox{\rm{[O\,\scalebox{.8}{I}]}}\xspace}
\def\oiii{\mbox{\rm{[O\,\scalebox{.8}{III}]}}\xspace}
\def\h2{\mbox{{\sc H}$_2$}\xspace}
\def\hi{\mbox{\rm{H\,\scalebox{.8}{I}}}\xspace}
\def\hii{\mbox{\rm{H\,\scalebox{.8}{II}}}\xspace}
\def\cii{\mbox{\rm{[C\,\scalebox{.8}{II}]}}\xspace}
\def\cplus{\mbox{\rm{C}$^{\rm{+}}$}\xspace}

\def\mgmc{\mbox{$m_{\rm{GMC}}$}\xspace}
\def\Mgmc{\mbox{$M_{\rm{GMC}}$}\xspace}
\def\Mgas{\mbox{$M_{\rm{gas}}$}\xspace}
\def\Mstar{\mbox{$M_{\rm{\ast}}$}\xspace}
\def\Mism{\mbox{$M_{\rm{ISM}}$}\xspace}

\def\Mdig{\mbox{$M_{\rm{diffuse\,ionized}}$}\xspace}
\def\Mdng{\mbox{$M_{\rm{diffuse\,neutral}}$}\xspace}

\def\nex{\mbox{$n_{\rm{H,ext}}$}\xspace}
\def\nH{\mbox{$n_{\rm{H}}$}\xspace}

\def\Tk{\mbox{$T_{\rm{k}}$}\xspace}

\def\fH2{\mbox{$f_{\rm{H}_2}$}\xspace}

\def\Pext{\mbox{P$_{\rm{ext}}$}\xspace}

\def\Pu{\mbox{\rm{K}\,\rm{cm}$^{-3}$}\xspace}
\def\Rgal{\mbox{R$_{\rm{gal}}$}\xspace}

\def\FUVMW{\mbox{$G_{\rm 0,MW}$}\xspace}

\def\d0{\mbox{$D_0$}\xspace}

\def\criMW{\mbox{$\zeta_{\rm{CR,MW}}$}\xspace}

\def\Z{\mbox{$Z$}\xspace}
\def\Zsfr{\mbox{$\langle Z\rangle_{\rm{SFR}}$}\xspace}
\def\Zmw{\mbox{$\langle Z\rangle_{\rm{mass}}$}\xspace}
\def\R{\mbox{$R$}\xspace}

\def\SFR{\mbox{\rm{SFR}}\xspace}

\def\SFRsd{\mbox{$\Sigma_{\rm{SFR}}$}\xspace}

\def\rh2{\mbox{$R_{\rm{H}_2}$}\xspace}
\def\rrh2{\mbox{$r_{\rm{H}_2}$}\xspace}

\def\rgmc{\mbox{$R_{\rm{GMC}}$}\xspace}

\def\Lcii{\mbox{$L_{\rm{[C\,\scalebox{.6}{II}]}}$}\xspace}

\def\Zsun{\mbox{$Z_\odot$}\xspace}
\def\lsun{\mbox{${\rm L_\odot}$}\xspace}
\def\cmpc{\mbox{cm$^{-3}$}\xspace} 
\def\kms{\mbox{km\,s$^{-1}$}\xspace}
\def\ps{\mbox{s$^{-1}$}\xspace}

\def\msun{\mbox{$\rm{M}_\odot$}\xspace}
\def\sfru{\mbox{$\rm{M}_\odot$~yr$^{-1}$}\xspace}

\def\sigame{\texttt{S\'IGAME}\xspace}
\def\cloudy{\mbox{\rm{C\scalebox{.8}{LOUDY}}}\xspace}

\def\gizmo{\texttt{GIZMO}\xspace}
\def\caesar{\texttt{CAESAR}\xspace}

\def\starburst{\texttt{starburst99}\xspace}
\def\powderday{\texttt{powderday}\xspace}
\newcommand{\music}{\mbox{\sc music}}
\newcommand{\mufasa}{\mbox{\sc mufasa}}

\newcommand{\e}[1]{\ensuremath{\times10^{#1}}}

\usepackage{color}
\usepackage{textpos}

\shorttitle{\sigame simulations of $z\simeq 6$ galaxies}
\shortauthors{Olsen et al.}

\begin{document}

\title{\sigame simulations of the \cii, \oi and \oiii line emission from star forming galaxies at $z\simeq 6$}

\correspondingauthor{Karen Olsen}
\email{kpolsen@asu.edu}

\author{Karen Olsen} 
\affil{School of Earth and Space Exploration, Arizona State University,
781 S Terrace Rd, Tempe, AZ 85287, USA}

\author{Thomas R. Greve} 
\affil{Department of Physics and Astronomy, University College London, 
Gower Street, London WC1E 6BT, UK}

\author{Desika Narayanan} 
\affil{Department of Astronomy, University of Florida, 211 Bryant Space Sciences Center, Gainesville, FL, USA}

\author{Robert Thompson} 
\affil{Portalarium, Austin, TX 78731, USA}

\author{Romeel Dav\'{e}} 
\affil{University of the Western Cape, 
Bellville, Cape Town 7535, South Africa}
\affil{South African Astronomical Observatory, 
Observatory, Cape Town 7925, South Africa}
\affil{Institute for Astronomy, Royal Observatory, 
Edinburgh EH9 3HJ, UK}

\author{Luis Niebla Rios} 
\affil{School of Earth and Space Exploration, Arizona State University, 781 S Terrace Rd, Tempe, AZ 85287, USA}

\author{Stephanie Stawinski} 
\affil{School of Earth and Space Exploration, Arizona State University, 781 S Terrace Rd, Tempe, AZ 85287, USA}



\begin{abstract}
Of the almost 40 star forming galaxies at $z\gs 5$ (not counting QSOs) observed in \cii to date, nearly
half are either very faint in \cii, or not detected at all, and fall well below expectations based on locally derived
relations between star formation rate and \cii luminosity.
This has raised questions as to how reliable \cii is as a tracer of star formation
activity at these epochs and how factors such as metallicity might affect the \cii emission. 
Combining cosmological zoom simulations of galaxies 
with \sigame (SImulator of GAlaxy Millimeter/submillimeter Emission) we have modeled the multi-phased interstellar medium (ISM) and its emission in \cii, as well as \oi and \oiii, from 30 main sequence galaxies at $z\simeq 6$ with star formation rates $\sim 3-23$\,\sfru, stellar masses $\sim(0.7-8)\times10^9\,{\rm \msun}$, and metallicities $\sim(0.1-0.4)\times\Zsun$. 
The simulations are able to reproduce 
the aforementioned \cii-faintness of some normal star forming galaxies sources at $z\ge 5$. 
In terms of \oi and \oiii very few observations are available at $z\gs 5$ -- but our simulations match two of the three existing $z\gs 5$ detections of \oiii, and are furthermore roughly consistent with the \oi and \oiii luminosity relations with star formation rate observed for local starburst galaxies.
We find that the \cii emission is dominated by the diffuse ionized gas phase and molecular clouds, which on average contribute $\sim 66\%$ and $\sim 27\%$, respectively. The molecular gas,
which constitutes only $\sim 10\%$   of the total gas mass is thus a more efficient emitter of \cii than the ionized gas, which makes up $\sim 85\%$ of the total gas mass.
A principal component analysis shows that the \cii luminosity correlates  with the star formation activity of a galaxy as well as its average metallicity.  
The low metallicities of our simulations together with their low molecular gas mass fractions can account for their \cii-faintness, and we suggest these  factors may also be responsible for the \cii-faint normal galaxies observed at these early epochs.
\end{abstract}

\keywords{cosmology: theory -- galaxies: high-redshift -- galaxies: ISM -- line: formation -- methods: numerical -- submillimeter: ISM}

\section{Introduction} \label{sec:intro}

The far-IR fine-structure transitions [C{\sc ii}] $^2P_{3/2} - ^2P_{1/2}$ at $157.7\,{\rm \mu m}$, [O{\sc i}] $^3P_{2} - ^3P_{1}$ at $63.2\,{\rm \mu m}$, and [O{\sc iii}] $^3P_{1} - ^3P_{0}$ at $88.4\,{\rm \mu m}$ (hereafter referred to as \cii, \oi, and \oiii) have been used as diagnostic tracers of the interstellar medium (ISM) and star formation activity for over two decades \citep[e.g.,][]{malhotra97,luhman98,fischer99,luhman03,sturm10,stacey10, ferkinhoff10, wang13a,delooze14,pineda14,sargsyan14,rigopoulou14,capak15,gullberg15,willott15,diaz-santos13,diaz-santos17}.
\cii is often observed to be one of the strongest emission lines in the spectra of galaxies, and can comprise up to $\sim 0.1-1\%$ of the infrared (IR) luminosity \citep{stacey91,helou01}. With C$^{\rm 0}$ having an ionization potential of only $11.3\,{\rm eV}$ and with the \cii line having an upper state energy of $E_{\rm u}/k_{\rm B} \sim 91\,{\rm K}$, several ISM phases can contribute to its emission. It is an important coolant in diffuse H{\sc i} clouds, diffuse ionized gas, and even molecular gas, where its critical density spans a wide range from $\sim 5\,{\rm cm^{-3}}$ for collisions with electrons at $\Tk=8000$\,K to $\sim 7.6\times 10^3\,{\rm cm^{-3}}$ for collisions with molecules at $\Tk=20$\,K \citep{goldsmith12}. In photo-dissociation regions (PDR), it is associated with both the interface layer of atomic gas, as well as from the ionized gas in the H{\sc ii} region itself \citep[e.g.,][]{stacey91,malhotra01,brauher08, smith17}.  \oi has a critical density of $\sim 4.7\times 10^5\,{\rm cm^{-3}}$ and an upper state energy of $E_{\rm u}/k_{\rm B} \sim 228\,{\rm K}$, which makes it an efficient tracer of PDRs. Since the ionization potential of O$^+$ is $35.1\,{\rm eV}$, \oiii is seen in hot diffuse ionized gas (e.g., H{\sc ii} regions or the hot ionized medium), where the transition is easily excited ($n_{\rm cr} \sim 5.1\times 10^2\,{\rm cm^{-3}}$ and $E_{\rm u}/k_{\rm B} \sim163\,{\rm K}$).

Although they trace different ISM phases, all three fine-structure lines have been proposed as tracers of the star formation rate (SFR) of galaxies \citep{Kapala15,herrera-camus15}. For example, \cite{delooze14} found the line emissions to correlate with the SFRs for a sample of local and high-$z$ starburst galaxies. However, the same study also found that for local metal-poor dwarfs, \oi and \oiii were better at predicting the SFR than \cii, which showed an increased scatter and a somewhat shallower correlation \citep{delooze14}.  Furthermore, local (ultra) luminous IR galaxies ((U)LIRGs) show a significant decrement in their \cii luminosity \citep[e.g.,][]{malhotra01}.
Self-absorption of the \oi line due to intervening cold or subthermally excited gas 
has also been reported in (U)LIRGS \citep[e.g.][]{rosenberg15} possibly decreasing the line strength and its ability to trace the star formation.

At redshifts $>5$, \cii has been detected in little more than a dozen normal star forming galaxies \citep{capak15,gullberg15,willott15,knudsen16,pentericci16,bradac17,decarli17,smit17}, and about the same number of non-detections (i.e., upper limits).  The non-detections, and a small number of detections, predominantly have very low \Lcii/SFR ratios and do not appear to form a \cii-SFR sequence with the remaining \cii-detected galaxies \citep{kanekar13,ouchi13,gonzalez-lopez14,ota14,schaerer15,maiolino15,inoue16}.
Proposed explanations for the low \cii emission include: low metallicities and thus C abundance; strong stellar feedback disrupting the neutral ISM from which most \cii emission is expected to arise \citep{maiolino15,inoue16,pentericci16,bradac17}; extreme UV-fields and thus a high ionization parameter $\langle U\rangle$\footnote{Number of far-UV photons per hydrogen atom} \citep{willott15}. 
An increase in $\langle U\rangle$ leads to higher grain charges (and grain temperatures) 
causing a deficiency of the photo-electrons available to heat the gas and hence 
a reduced photoelectric heating efficiency. 
Additionally, the effects of stellar age have been invoked \citep{gonzalez-lopez14,schaerer15} and different (possibly much denser) photodissociation region (PDR) structures than seen locally \citep{ouchi13,ota14}.

With the low number of \oi and \oiii detections at high redshift, it is practically impossible to establish an \oi-SFR or \oiii-SFR relation for the high redshift ($0.5<z<6.6$) sample of \cite{delooze14}. Since that study, two Ly$\alpha$ emitters (LAEs) at $z=7.2120$ and $8.38$, and a normal star forming galaxy at $z=7.1$ have been detected in \oiii \citep{inoue16,laporte17,carniani17}. 

\smallskip

\smallskip

Galaxy-scale simulations have been developed that attempt to account for the observations of the above fine-structure lines, in particular, how their emissions are linked to the ISM phases and to global galaxy properties, such as metallicity, star formation efficiency, ionization parameter, dust mass fraction, compactness and phase filling factors \citep[e.g.,][]{cormier12,vallini13,olsen15,accurso17,katz17}. By combining codes of stellar population synthesis, radiative transfer, photoionization, and astrochemistry into simulations of starburst regions, \cite{accurso17} found that the increases in the specific star formation rate of a galaxy leads to a decrease in the fraction of the \cii emission coming from the molecular gas phase, due to stronger UV radiation fields which will tend to shrink the molecular regions.  Applied to local normal galaxies, their code predicts that as much as $60-80\%$ of the \cii emission emerges from the molecular gas. \cite{cormier12} utilized the photoionisation code \cloudy \citep{ferland13} to model a number of FIR fine-structure line emissions (including \cii, \oi, and \oiii) from the multi-phased ISM in the starburst low-metallicity galaxy Haro 11. They found that PDRs account for only 10\,\% of the \cii emission, with the remaining emission arrising in the diffuse ionized medium, but a larger PDR contribution when lowering the density or including magnetic fields.

Some simulation studies have focused specifically on $z \gs 6$ galaxies using either cosmological simulations \citep{vallini13,vallini15,pallottini17} or an analytical framework \citep{munoz14,bonato14,popping16,narayanan17}. Simulating galaxies at $z\approx6.6$, \cite{vallini15} parametrises their \cii luminosities as a function of SFR and metallicity.  They find that the \cii-faint sources at these epochs are explained by either low metallicity or negative feedback from regions of intense star formation. They also find that the \cii budget is dominated by emission from PDRs, with $<10\%$ coming from the diffuse neutral gas. \cite{pallottini17} calculated the \cii emission from a multi-phase ISM in a normal star forming galaxy at $z\simeq 6$ by combining abundance calculations using \cloudy with the method of inferring the \cii emission by \cite{vallini13,vallini15} and applying it to a cosmological simulation. They too were able to reproduce the underluminous \cii sources found at this epoch and, just like \citet{vallini15} attributed it to low metallicity. They also find that while $\sim 95\,\%$ of the total \cii luminosity is coming from the main gas disk of the galaxy, about $30\%$ of the total C$^+$ mass is situated in an outflow. Owing to the low density of the outflow, however, this mass does not contribute significantly to the total \cii luminosity.

In this paper we simulate the \cii, \oi and \oiii line emission from normal star forming galaxies at $z\simeq6$, although the emphasis of our analysis will be on \cii. Due to the different origins of the lines, a multi-phased modelling of the ISM is required.  To this end we have combined the output from cosmological zoom simulations of normal star forming galaxies at this epoch with an updated version of \sigame \citep[S\'Imulator of GAlaxy Millimeter/submillimeter Emission;][]{olsen15}. \S\ref{SPH} describes the simulation codes used for the galaxy evolution and the galaxy sample chosen for our study.  The sub-grid modelling of the ISM is done with \sigame and is described in \S\ref{MOD}.  The simulation results are presented, analyzed and discussed in \S\ref{results} and \S\ref{dis}, respectively, followed by our conclusions in \S\ref{con}. Throughout, we adopt a flat cold dark matter ($\Lambda$CDM) cosmology with cosmological parameters; $\Omega_{\Lambda}=0.7$, $\Omega_{\rm M}=0.3$ and $h=0.68$ \citep{planck16}.

\section{Cosmological Galaxy Formation simulations}\label{SPH}

We use cosmological zoom simulations of galaxies
extracted from the \mufasa\ cosmological simulation \citep{dave16a,dave17},
utilizing the same feedback prescription as in \mufasa.
We first briefly summarize the zoom technique, then discuss the
particulars of the \mufasa\ simulations.

The zoom technique is summarized in recent reviews
\citep[e.g.][]{somerville15b}, hence we only condense the salient
details here.
We begin by simulating a dark matter-only $50
h^{-1}$\,Mpc$^3$ volume utilizing
\gizmo\footnote{\url{http://www.tapir.caltech.edu/~phopkins/Site/GIZMO.html}},
which employs the {\sc Gadget-2} tree-particle-mesh gravity solver~\citep{springel05}.  
This is run at fairly 
low resolution, with $512^3$ particles resulting in a particle mass of $m_{\rm DM} = 7.8
\times 10^{8} h^{-1}\,M_\odot$, from $z=249$ to $z=0$ with the aim of
simulating a population of dark matter halos to be re-simulated (with
baryons) at much higher mass resolution.  The initial conditions are
generated with \music \ \citep{hahn11a}, and have the exact same
random seeds as the cosmological \mufasa \ simulation with gas dynamics.  

At $z=2$, we identify a randomly-selected set of halos within the mass range 
$\sim(0.4-3)\times10^{13}$\,\msun for re-simulation at higher resolution, now including baryons.
We build an elliptical mask around each halo extending to $2.5 \times$
the distance of the farthest dark matter particle from the center of
the halo, and define this as the Lagrangian high-resolution region to
be re-simulated.  We seed this region with higher-resolution dark
matter ($m_{\rm DM}=1\times10^{6}$ h$^{-1}$\,M$_\odot$) and gas
($1.9\times10^{5}$\,h$^{-1}$\,M$_\odot$) fluid elements, and resimulate.   We employ adaptive gravitational
softening of all particles throughout the simulation
\citep{hopkins15}, with minimum force softening lengths of $12, 3$
and $3$ pc for dark matter, gas and stars respectively. 

The hydrodynamic simulations are run with a modified version of the
$N$-body/hydrodynamics solver, \gizmo \ \citep{hopkins15} using
the meshless finite mass (MFM) hydrodynamics solver.  MFM uses a
Riemann solver to compute pressure gradients, whose solution is
obtained in a frame chosen to conserve mass within each mesh cell.
This retains essential advantages of mesh methods that handle shocks
and contact discontinuities accurately, along with that of particle-based
methods in which mass can be tracked and followed.

These simulations utilize the suite of physics developed for the
\mufasa \ cosmological simulations as described in 
\citet{dave16a,dave17}.  We refer the reader to these papers for more detail, and
summarize the relevant details here.  Star formation occurs only in
dense molecular gas, where the ${\rm H_2}$ gas fraction (\fH2) is determined
based on the \citet{krumholz09} and \citet{krumholz11} relations that
approximate the molecular gas mass fraction as a function of the gas phase
metallicity and surface density.  Due to our high resolution, we assume no
clumping on scales below our resolution limit.  Star formation follows
a volumetric \citet{schmidt59} relation with an imposed star formation
efficiency of $\sigma_* = 0.02$.

Young stars generate winds using a decoupled, two-phase wind.  These
winds have a probability for ejection that is a fraction of the star
formation rate probability, according to the best-fit
relation from the Feedback in Realistic Environments simulations
studied (FIRE) by \citet{muratov15a} \ \citep[and is additionally motivated 
by analytic arguments;][]{hayward15a}.  The ejection
velocity scales with the galaxy circular velocity again following the
\citet{muratov15a} scaling relations.  These winds are decoupled from
hydrodynamic forces or cooling until either its relative velocity to
the background gas is less than half the local sound speed; the wind
reaches a density limit less than $0.01 \times $ that of the critical
density of star formation, or at least $2\%$ of the Hubble time 
has elapsed since the time of launch.

 \begin{figure}[ht]
 \begin{center}
 \epsscale{1.2}
 \includegraphics[width=\columnwidth]{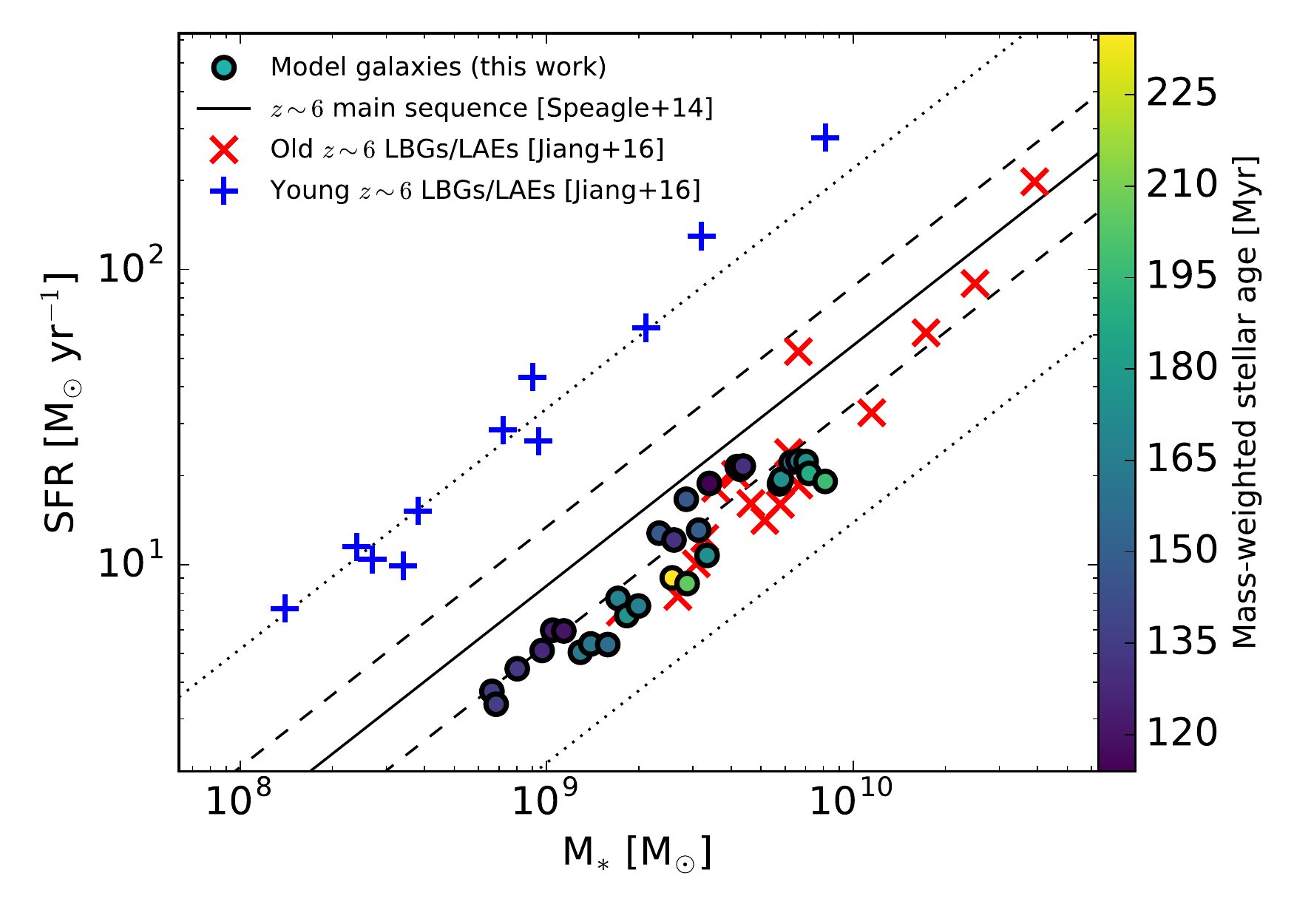}
 \end{center}
 \caption{SFR vs.\ $M_{\rm \ast}$ for our \nsam simulated galaxies 
 (filled circles, color-coded according to their stellar mass weighted average ages). 
 The observed ${\rm SFR}-M_{\rm \ast}$ relation at $z\simeq 6$ as determined by
 \citet{speagle14} is shown as a solid line 
 accompanied by dashed and dotted lines indicating the
 $1\sigma$ and $3\sigma$ scatter around the relation, respectively. For comparison we show the positions of $z\sim6$ "old" ($>100\,$Myr) and "young" ($<30\,$Myr) LBGs/LAEs \cite{jiang16}.
}
 \label{M_SFR}
 \end{figure}

Feedback from longer-lived stars (i.e. Asymptotic Giant Branch stars
and Type Ia supernovae) is also included.  These delayed feedback
sources follow \citet{bruzual03a} tracks with a \citet{chabrier03}
initial mass function, and deposit heavy metals (H, He, C, N, O, Ne,
Mg, Si, S, Ca, and Fe) into the interstellar medium.  The chemical
enrichment yields for Type II SNe, Type Ia SNe and AGB stars are taken
from \citet{nomoto06a,iwamoto99a} and \citet{oppenheimer08}
parameterisations respectively.

We extract model galaxies from snapshots at $5.875<z<6.125$ with
stellar masses (\Mstar) and SFRs in the range $\sim (0.7-8)\times 10^9\,{\rm \msun}$ and $\sim 3-23\,{\rm \msun\,yr^{-1}}$, respectively.
In total, we select \nsam galaxies, hereafter named G1, ..., G\nsam by
order of increasing stellar mass.  Fig.\,\ref{M_SFR} shows the
locations of G1, ..., G\nsam in the SFR$-$\Mstar diagram.  The
galaxies are consistent with the observational determination of the
$z\sim6$ main sequence (MS) of star forming galaxies by
\cite{speagle14}, although slightly offset towards higher \Mstar for a given SFR. Also shown in Fig.\,\ref{M_SFR} are the $z\sim6$ LBGs and LAEs studied by \cite{jiang16} who divided their sample into "old" ($>100\,$Myr) and "young" ($<30\,$Myr) galaxies, depending upon their stellar age as derived from spectral energy distribution (SED) fitting. As the color-coding shows, our model galaxies are consistent in terms of \Mstar, SFR and stellar age with the "old" LBG/LAE population \citep[][]{jiang16}.
Table \ref{table:1} gives an overview of other
global parameters of our simulations that are relevant for this study.  The SFR of each galaxy was
calculated as a mean over the past 100\,Myr (mass of new stars created
over the past 100\,Myr divided by 100\,Myr) and the radius of the disk is 
calculated by the analysis software
\caesar\footnote{\url{https://bitbucket.org/rthompson/caesar}}, an extension of the {\sc yt} simulation tools\footnote{\url{http://yt-project.org/}},
using a friends-of-friends algorithm.
Table \ref{table:1} also lists the 
average metallicity of each simulated galaxy, which is calculated as the SFR-weighted average over all the fluid elements. These SFR-weighted average metallicities, \Zsfr,
range from $14\%$ to $45\%$ of the solar metallicity. 

\begin{deluxetable*}{c|cccccccccc}[ht]
\tablecaption{Global properties of the \nsam simulated $z\simeq6$ galaxies used for this work.\label{table:1}}
\tablewidth{0pt}
\tablehead{
\colhead{Name} & \colhead{$z$} & \colhead{\Rgal\tablenotemark{a} [kpc]} & \colhead{\Mstar [10$^{9}$\,\msun]} & \colhead{\Mgas [10$^{9}$\,\msun]} & \colhead{SFR\tablenotemark{b} [\sfru]} & \colhead{\Zsfr\tablenotemark{c} [\Zsun]} & \colhead{$\frac{\Mgmc}{\Mgas}$} & \colhead{$\frac{\Mdng}{\Mgas}$} & \colhead{$\frac{\Mdig}{\Mgas}$}
}
\startdata
G1 & 	6.12 & 	9.34 	& 	0.663 & 	11.493 & 	3.720 & 	0.186 & 	0.071 & 	0.059 & 	0.869 \\
G2 & 	6.25 & 	7.45 	& 	0.684 & 	8.170 & 	3.365 & 	0.136 & 	0.093 & 	0.063 & 	0.844 \\
G3 & 	6.00 & 	8.12 	& 	0.802 & 	11.810 & 	4.437 & 	0.178 & 	0.089 & 	0.077 & 	0.833 \\
G4 & 	5.88 & 	9.67 	& 	0.967 & 	11.821 & 	5.118 & 	0.183 & 	0.075 & 	0.054 & 	0.871 \\
G5 & 	5.88 & 	8.86 	& 	1.047 & 	13.016 & 	5.991 & 	0.170 & 	0.140 & 	0.099 & 	0.761 \\
G6 & 	5.75 & 	7.49 	& 	1.138 & 	10.259 & 	5.959 & 	0.152 & 	0.115 & 	0.090 & 	0.794 \\
G7 & 	6.00 & 	9.25 	& 	1.289 & 	11.499 & 	5.045 & 	0.156 & 	0.089 & 	0.079 & 	0.831 \\
G8 & 	6.12 & 	11.29 	& 	1.393 & 	14.489 & 	5.384 & 	0.155 & 	0.088 & 	0.064 & 	0.847 \\
G9 & 	6.25 & 	8.31 	& 	1.583 & 	9.170 & 	5.358 & 	0.198 & 	0.143 & 	0.059 & 	0.797 \\
G10 & 	6.25 & 	5.70 	& 	1.710 & 	9.532 & 	7.681 & 	0.159 & 	0.216 & 	0.030 & 	0.754 \\
G11 & 	5.75 & 	12.68 	& 	1.826 & 	16.511 & 	6.719 & 	0.236 & 	0.063 & 	0.054 & 	0.882 \\
G12 & 	5.88 & 	11.37 	& 	1.994 & 	15.293 & 	7.239 & 	0.211 & 	0.020 & 	0.119 & 	0.861 \\
G13 & 	6.12 & 	7.41 	& 	2.329 & 	10.718 & 	12.776 & 	0.208 & 	0.107 & 	0.022 & 	0.871 \\
G14 & 	5.75 & 	11.05 	& 	2.569 & 	14.296 & 	9.014 & 	0.231 & 	0.103 & 	0.062 & 	0.835 \\
G15 & 	6.00 & 	9.07 	& 	2.603 & 	12.520 & 	12.115 & 	0.242 & 	0.053 & 	0.107 & 	0.840 \\
G16 & 	6.00 & 	7.70 	& 	2.856 & 	11.051 & 	16.644 & 	0.202 & 	0.180 & 	0.022 & 	0.798 \\
G17 & 	5.88 & 	6.67 	& 	2.873 & 	8.119 & 	8.625 & 	0.258 & 	0.214 & 	0.070 & 	0.716 \\
G18 & 	5.75 & 	13.84 	& 	3.134 & 	19.713 & 	13.078 & 	0.217 & 	0.020 & 	0.090 & 	0.889 \\
G19 & 	6.12 & 	10.46 	& 	3.338 & 	10.771 & 	10.739 & 	0.282 & 	0.023 & 	0.129 & 	0.848 \\
G20 & 	6.25 & 	7.78 	& 	3.395 & 	10.299 & 	18.833 & 	0.334 & 	0.021 & 	0.025 & 	0.954 \\
G21 & 	6.00 & 	4.15 	& 	4.186 & 	7.822 & 	21.474 & 	0.412 & 	0.072 & 	0.032 & 	0.896 \\
G22 & 	6.12 & 	6.40 	& 	4.277 & 	9.044 & 	21.042 & 	0.452 & 	0.020 & 	0.033 & 	0.947 \\
G23 & 	6.25 & 	6.70 	& 	4.381 & 	10.061 & 	21.558 & 	0.339 & 	0.031 & 	0.028 & 	0.941 \\
G24 & 	6.25 & 	7.47 	& 	5.750 & 	11.261 & 	18.776 & 	0.293 & 	0.157 & 	0.014 & 	0.828 \\
G25 & 	5.88 & 	7.03 	& 	5.833 & 	9.567 & 	19.442 & 	0.354 & 	0.039 & 	0.038 & 	0.923 \\
G26 & 	6.12 & 	10.43 	& 	6.278 & 	14.223 & 	22.182 & 	0.371 & 	0.138 & 	0.056 & 	0.805 \\
G27 & 	6.00 & 	10.84 	& 	6.622 & 	13.159 & 	22.455 & 	0.370 & 	0.103 & 	0.052 & 	0.844 \\
G28 & 	5.88 & 	10.52 	& 	7.024 & 	9.680 & 	22.332 & 	0.425 & 	0.139 & 	0.046 & 	0.815 \\
G29 & 	5.75 & 	7.38 	& 	7.178 & 	6.710 & 	20.364 & 	0.392 & 	0.201 & 	0.013 & 	0.785 \\
G30 & 	5.75 & 	6.48 	& 	8.104 & 	9.095 & 	19.128 & 	0.409 & 	0.039 & 	0.030 & 	0.931 \\
\enddata
\tablenotetext{a}{\Rgal is the radius defined by the group finder 
\caesar together with the yt project (see text).}
\tablenotetext{b}{The SFR has been averaged over the past $100$\,Myr.}
\tablenotetext{c}{\Zsfr is the average of the SFR-weighted metallicities, in units of solar metallicity, of the fluid elements belonging to a galaxy.}
\end{deluxetable*}

\begin{figure*}[ht]
\epsscale{1.1}
\plotone{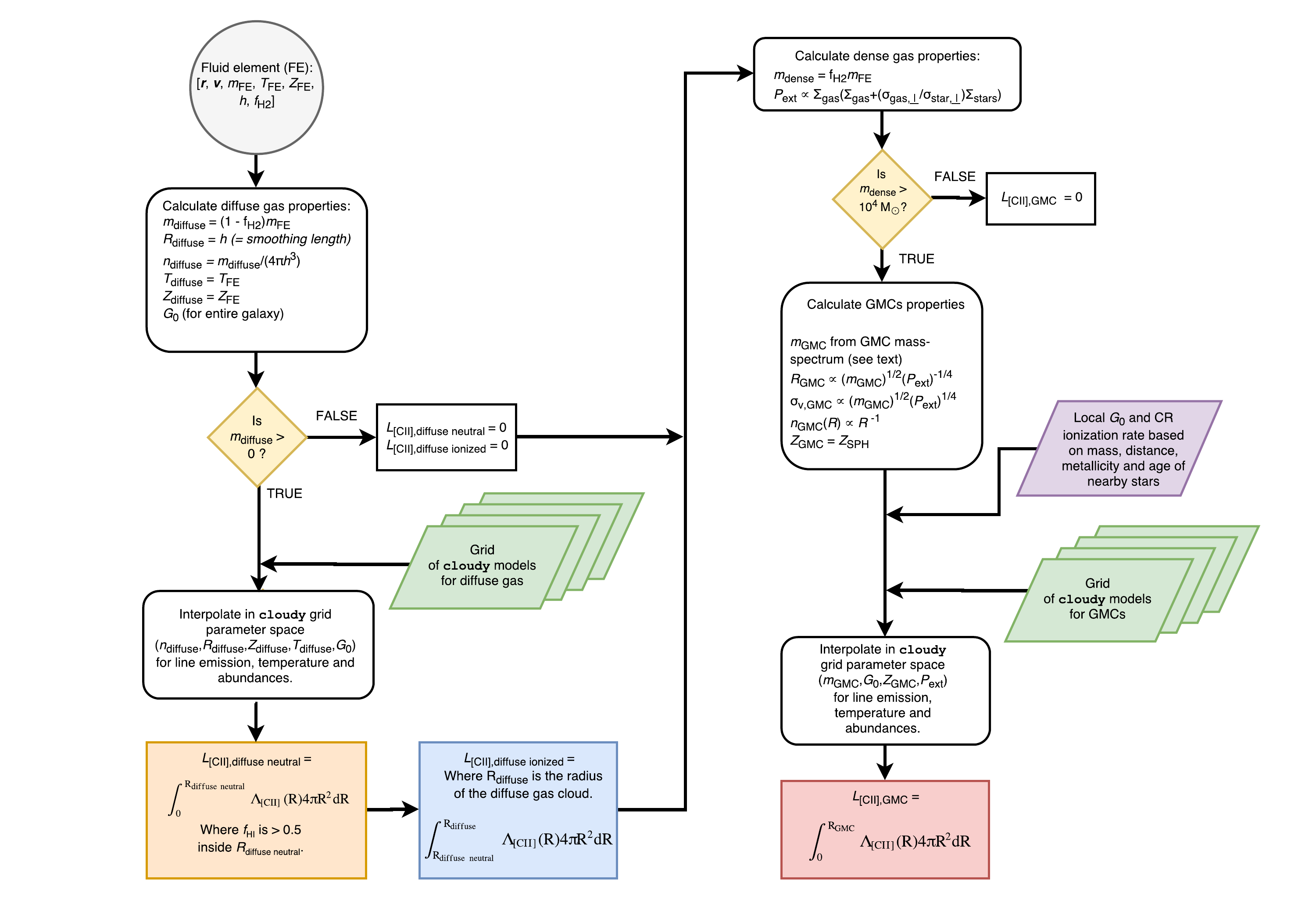}
\caption{Flowchart of the subgridding procedures applied by \sigame to each gas fluid element in the \mufasa~simulations. 
\label{overview}}
\end{figure*}

\section{\sigame}\label{MOD}
For the purposes of this paper we have updated \sigame from the version
presented in \citet{olsen15}. The main updates are: a more sophisticated 
calculation of the UV radiation fields, and the implementation of \cloudy\footnote{\url{http://www.nublado.org/}} ver. 17 \cite{ferland17}. Below we describe in detail the updates 
made.  Fig.\,\ref{overview}
illustrates how a gas fluid element is processed in this updated version
of \sigame.

\subsection{Dense phase}\label{MOD_dense}
\sigame uses a method described in \cite{rahmati13} to 
derive the dense gas mass fraction, \fH2, associated 
with a given fluid element. The dense gas mass is then
$m_{\rm dense}		=	m_{\rm gas}\fH2$,
where $m_{\rm gas}$ is the total gas mass of the fluid element. 
The dense gas is divided into giant molecular clouds (GMCs) by 
randomly sampling the Galactic GMC mass spectrum over the 
mass range $10^4-10^6\,\msun$ until the remaining dense gas 
mass is $< 10^4\,\msun$ at which point the gas is discarded. 
If the initial $m_{\rm dense}$ does not exceed 
this lower limit no GMCs are associated with the fluid element 
and $m_{\rm dense}$ is set to zero. No more than $0.01\,\%$ 
of the dense gas mass in any of our \nsam simulations are 
discarded during this process. The Galactic GMC mass spectrum adopted here
has a power law slope of $1.8$ as derived by \cite{blitz07} for the outer MW. In \S\ref{dis} we explore
the effects of adopting a shallower as well as a steeper spectrum.

 \begin{figure}[ht]
 \begin{center}
 \epsscale{1.29}
 \plotone{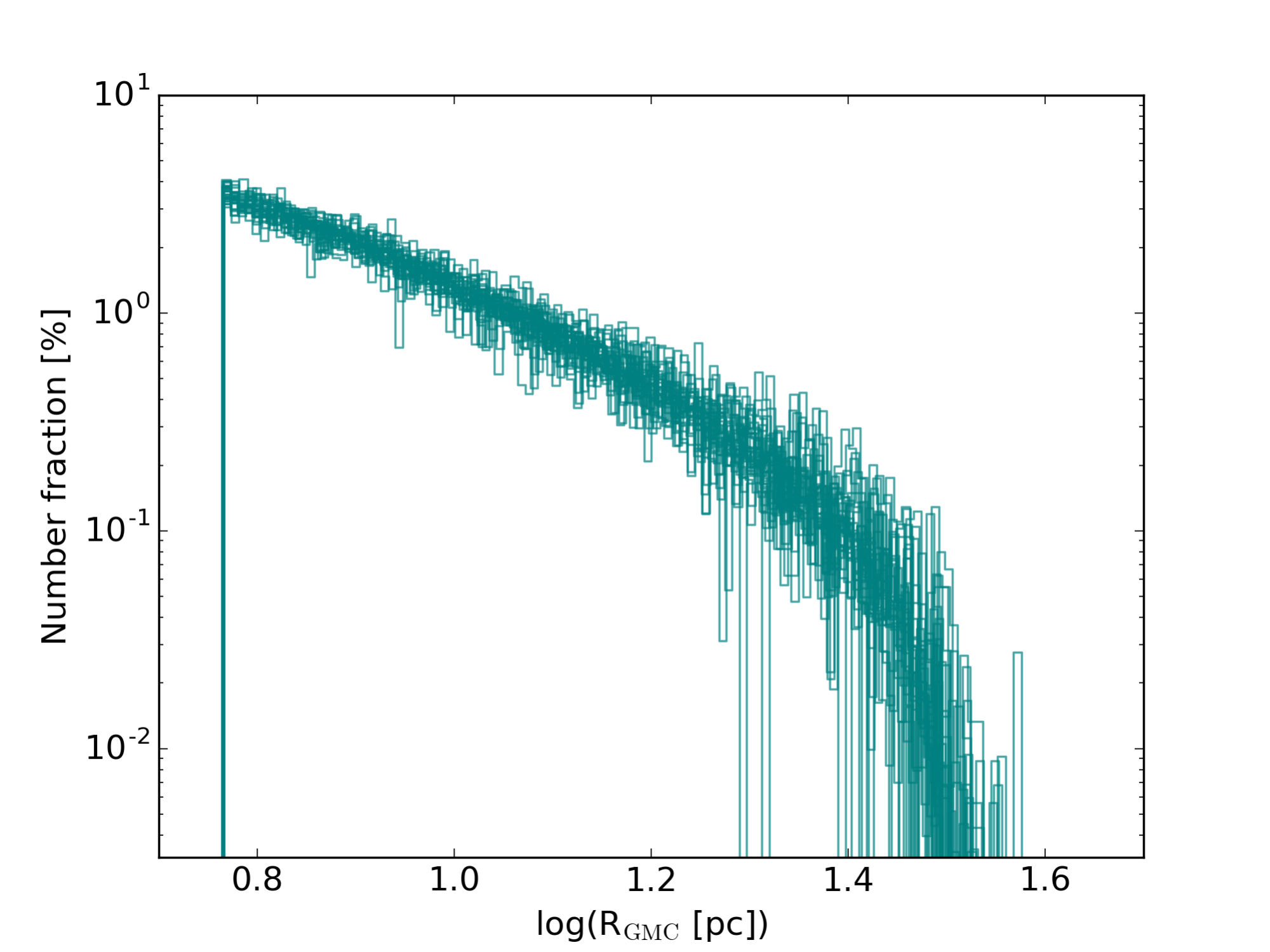}
 \end{center}
 \caption{The distributions of cloud radii for the GMC populations in our 30 model galaxies.\label{Rgmc}}
 \end{figure}

Each GMC inherits the metallicity, \Z, of its parent fluid element, and is subjected to an external cloud pressure, \Pext, which is calculated from the surface densities of stars and star-forming gas as well as their vertical velocity dispersions using an assumption of hydrostatic 
mid-plane equilibrium \citep[see][]{olsen15}. The GMC sizes ($R_{\rm GMC}$) are derived from a pressure-normalized mass-size relation \citep{olsen15}, which results in cloud radii in the range $5.8$ to $43.6$\,pc (Fig.\,\ref{Rgmc}).
The GMC radial density profiles are assumed to follow a truncated logotropic profile:
\begin{eqnarray}
	\nH(R)			=	\nex\left( \frac{\rgmc}{R} \right),	
	\label{eq:nH}
\end{eqnarray}
where $\nH(R>\rgmc)=0$ and the external density, \nex, is $2/3$ of the average density within \rgmc. 
The GMCs are randomly distributed within $0.5\times$ the
smoothing length of the fluid element. 

\bigskip

Each GMC is assumed to be isotropically irradiated by a local FUV radiation field
from nearby young stars and a constant diffuse background FUV field 
from the overall stellar population of the galaxy.
The local field is the cumulative FUV 
radiation field from stellar particles 
within one smoothing length of the gas fluid element position 
($\boldsymbol{r}_{\rm gas}$). Each contribution is 
scaled according to the stellar mass ($m_{*,i}$) and distance 
($|\boldsymbol{r}_{\rm gas}-\boldsymbol{r}_{{\rm *,}i}|$) 
from the gas element such that the strength of the local field is given by:
\begin{eqnarray}
\frac{G_{\rm 0, loc}}{{\rm erg\,cm^{-2}\,s^{-1}}} = \sum_{|\boldsymbol{r}_{\rm gas}-\boldsymbol{r}_{\rm *,i}| < h}^{} \frac{L_{{\rm FUV,}i}}{4\pi |\boldsymbol{r}_{\rm gas}-\boldsymbol{r}_{\rm *,i}|^2}\frac{m_{*,i}}{10^4\,\msun}
\label{FUV}
\end{eqnarray}
where $|\boldsymbol{r}_{\rm gas}-\bm{r}_{{\rm *,}i}|$ is in cm and $m_{*,i}$ is in \msun.  
$L_{{\rm FUV},i}$ is the FUV luminosity 
in units of ${\rm erg\,s^{-1}}$ of a $10^4\,\msun$ stellar 
population with a metallicity and age as that of the $i$th stellar 
particle.  The $L_{{\rm FUV},i}$-values are found by interpolating 
over a grid of
\starburst\footnote{\url{http://www.stsci.edu/science/starburst99/}}
\citep{leitherer14} stellar population models of mass $10^4\,\msun$  and
covering the range of metallicities and ages of the stellar populations
encountered in the simulations. 
These models adopt a Kroupa initial mass function (IMF) and an instantaneous burst star formation history.  
We note that the hydrodynamic simulations use a slightly different IMF 
(Chabrier), but expect the differences in UV luminosity to be negligible.
The strength of the constant background FUV field is set to:
\begin{equation}
\frac{G_{\rm 0, bg}}{{\rm erg\,cm^{-2}\,s^{-1}}} = G_{\rm 0,MW} \frac{\Sigma_{\rm SFR}}{\Sigma_{\rm SFR,MW}},
\label{G0-diffuse}
\end{equation}
where $G_{\rm 0,MW}=9.6\e{-4}$\,erg\,cm$^{-2}$\,s$^{-1}$ ($=0.6\,{\rm Habing}$) is the Galactic
FUV field flux \citep{seon11}, and $\Sigma_{\rm SFR}$ and $\Sigma_{\rm SFR,MW}$ are the average
SFR surface densities of the model galaxy in question and the MW, respectively. 
\SFRsd is defined as the total SFR divided by the area of the
galaxy disk seen face-on, using the radii listed in Table\,\ref{table:1}. 
Our model galaxies span a range in \SFRsd from 
$0.013$ to $0.40\,{\rm \sfru\,kpc^{-2}}$, which corresponds to $4-120\times \Sigma_{\rm SFR, MW}$ since $\Sigma_{\rm SFR, MW}\simeq 0.003\,\sfru$. The latter
is calculated using a SFR of $1.9$\,\sfru \citep{chomiuk11} and a radius of 13.5\,kpc 
for the star forming disk of the MW \citep{kennicutt12}. 

The total FUV flux impinging 
on a GMC is $G_{\rm 0, GMC} = G_{\rm 0, loc}+G_{\rm 0, bg}$. For the sake
of simplicity we assume that the spectral shape of this FUV field is identical to that
of the standard FUV background in the Solar neighborhood (the "ism" table in \cloudy).

\bigskip

Cosmic rays (CRs) can be an important source of heating and 
ionization in dense clouds \citep{papa11,papa14}, and are 
therefore included in the treatment of GMCs in \sigame.  
Each GMC is subjected to a CR ionization rate, which is
given by:
\begin{equation}
\frac{\zeta_{\rm CR, GMC}}{{\rm s^{-1}}} = \criMW \frac{G_{\rm 0, GMC}}{G_{\rm 0,MW}},
\end{equation}
where $\criMW=3\times10^{-17}\,\ps$ is the average CR
ionization rate in the MW \citep[e.g.,][]{webber98}.

\bigskip

\begin{figure*}[h]
\epsscale{0.9}
\plotone{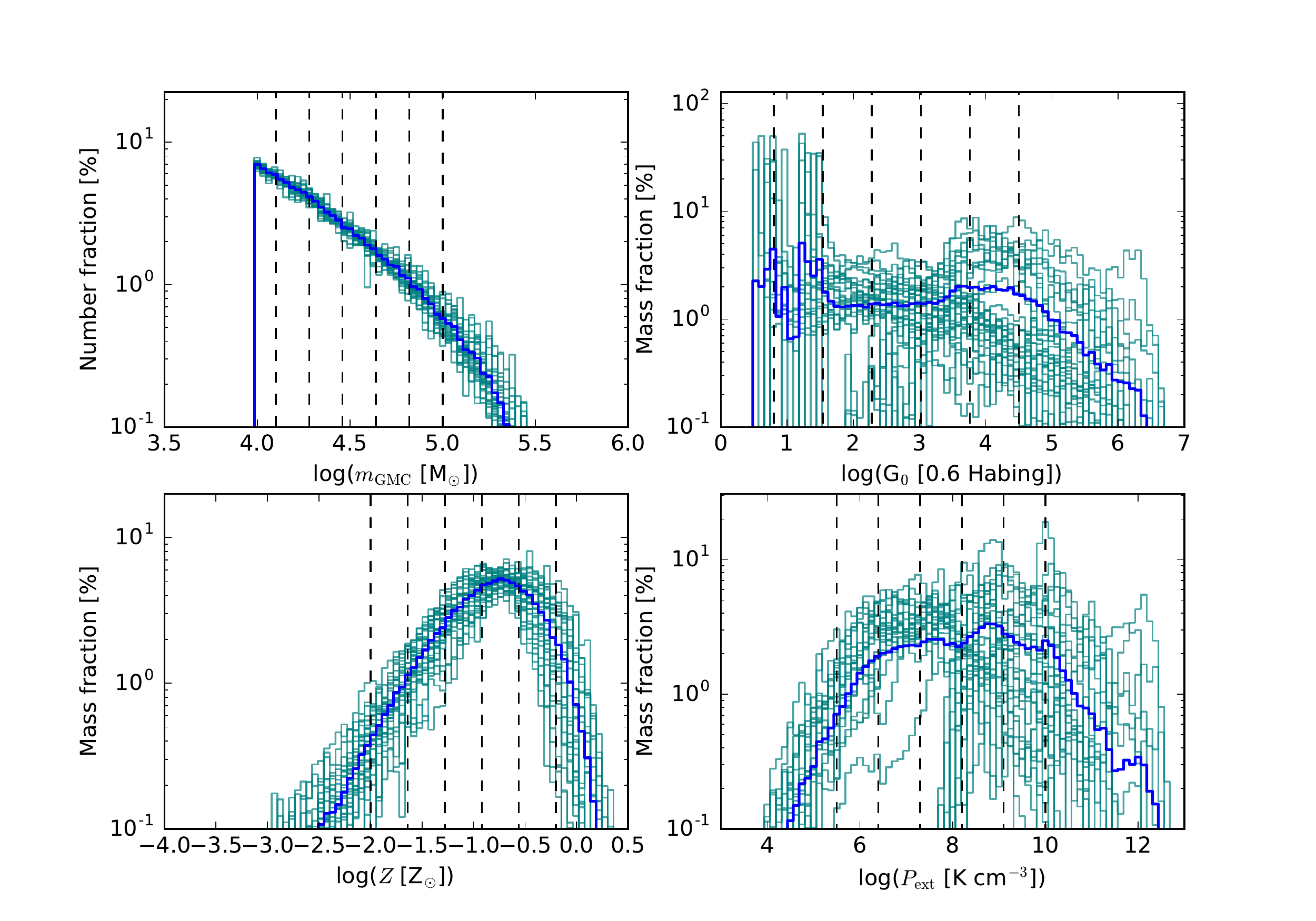}
\caption{Histograms of \mgmc, $G_{\rm 0, GMC}$, \Z, and \Pext of the GMCs in our \nsam model galaxies (dark green histograms), with the mean histograms shown in blue. 
The vertical dashed lines indicate the chosen \texttt{CLOUDY} grid points (see \S\ref{MOD_clo}).\label{GMCgridpoints}}
\end{figure*}

The distributions of \mgmc, $G_{\rm 0, GMC}$, \Z and \Pext for the 
GMC populations in our simulations are shown in Fig.\,\ref{GMCgridpoints}. 
Once the GMCs have been configured in the above described 
manner, \cloudy is used to calculate the ionization states, thermal state and 
line cooling rates throughout the clouds. 
The line luminosities of a GMC is calculated as the line cooling 
rates provided by \cloudy integrated over the volume of the 
cloud.

\subsection{Diffuse phase}\label{MOD_dif}
\begin{figure*}[h]
\epsscale{0.9}
\plotone{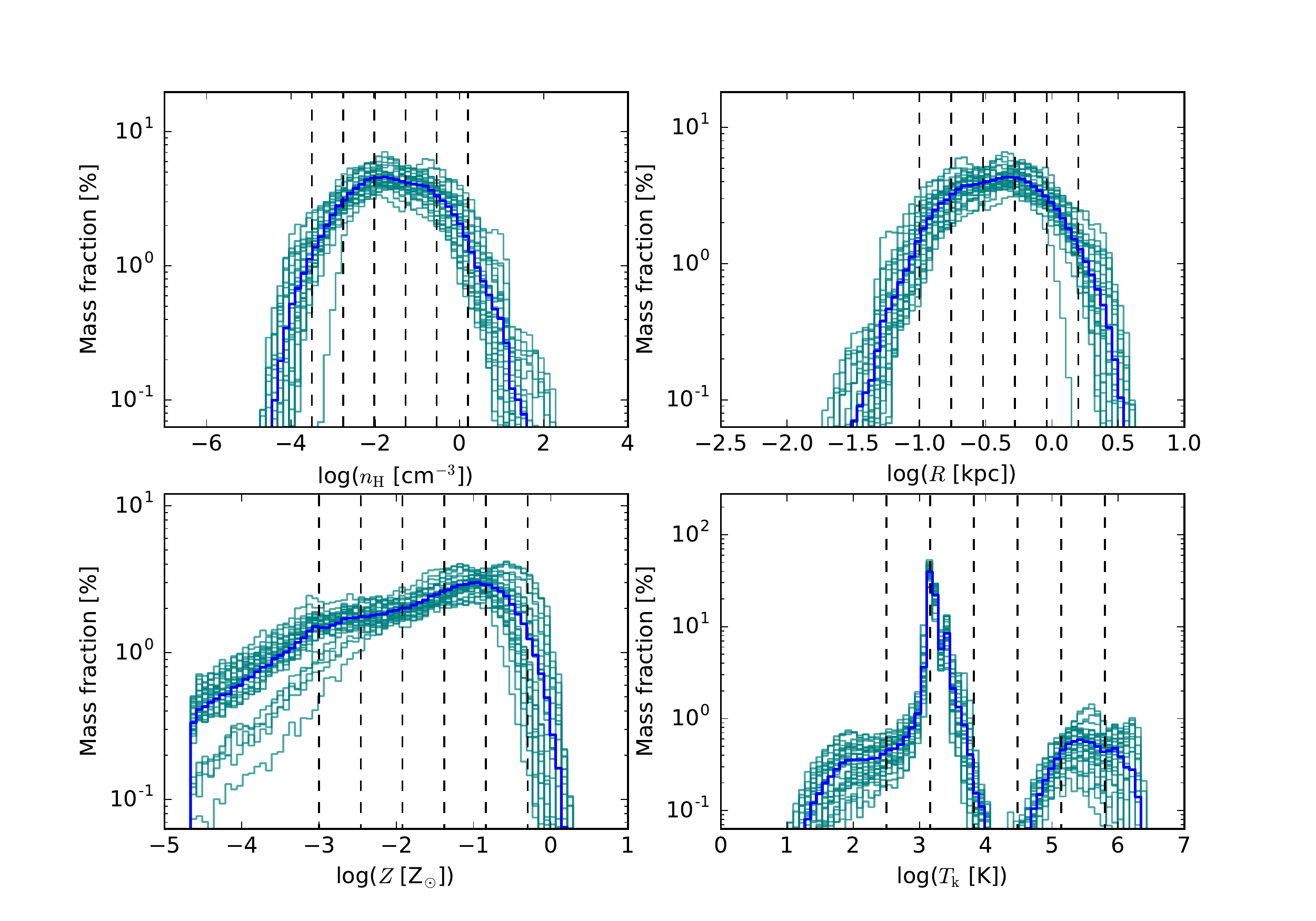}
\caption{Histograms of \nH, $R$, \Z, and \Tk of the diffuse gas clouds in our \nsam model galaxies (dark green histograms), with the mean histograms shown in blue. 
The vertical dashed lines indicate the chosen \texttt{CLOUDY} grid points (see \S\ref{MOD_clo})\label{difgridpoints}. 
}
\end{figure*}

The diffuse gas mass associated with a fluid element is 
$m_{\rm diffuse} = m_{\rm gas} - m_{\rm dense}$. 
Provided $m_{\rm diffuse} > 0$ for the fluid element in question, 
the diffuse gas is distributed evenly within a spherical region 
with a radius equal to the smoothing length of the fluid element and centered on its position. 
Furthermore, the diffuse gas inherits the metallicity and 
temperature of the fluid element. Fig.\ \ref{difgridpoints}
shows the distributions of densities, radii, metallicities, 
and temperatures for the diffuse clouds in our model galaxies.

\bigskip

The clouds of diffuse gas are larger 
(typical radii are $\sim 1\,{\rm kpc}$ but span the range $0.1-5\,{\rm kpc}$) than the
dense gas clouds and have lower densities (typical densities are $\sim 10^{-2}\,{\rm cm^{-3}}$ but span the range $\sim 10^{-4}- 10^2\,{\rm cm^{-3}}$).  
It is assumed that the diffuse clouds are unassociated with
star formation sites and, as a result, they are only irradiated (isotropically)
by the diffuse background FUV field (i.e., $G_{\rm 0,bg}$).
The heating and ionization of the diffuse gas is also affected by the presence of CRs, and
we set the CR ionization rate felt by the diffuse phase to:
\begin{equation}
\frac{\zeta_{\rm CR, bg}}{{\rm s^{-1}}} = \criMW \frac{G_{\rm 0, bg}}{G_{\rm 0,MW}}.
\end{equation}

\bigskip

As done for the GMCs, \cloudy is used to calculate the ionization states and line cooling rates
throughout the diffuse clouds. 
\sigame uses 
the output from \cloudy to define the transition from the inner 
neutral to the outer ionized regions of the diffuse clouds: the 
radius where the neutral hydrogen fraction,
i.e., the number density of H{\sc i} divided by the number density of H{\sc i} and H{\sc ii}, is 0.5. In cases where this fraction is
$> 0.5$ at all radii, \sigame identifies the entire cloud 
as diffuse neutral gas. The above procedure allows us to operate 
with a diffuse neutral and a diffuse ionized gas phase in our simulations, in addition to the
dense gas phase. The line luminosities of the neutral 
and ionized phases within a cloud are calculated by integrating the line cooling 
rates over the volumes of each of the 
two phases.

\subsection{CLOUDY model setup}\label{MOD_clo}
The main input parameters for \cloudy in this work are the cloud radius, the radial density profile of the cloud, the FUV spectrum 
impinging on the cloud, the element abundances, and the gas kinetic temperature. In the following we describe how each of these parameters are specified for GMCs and diffuse gas clouds. \\

\noindent {\it Cloud radius: } For our GMC models, the external pressure is  used together with the GMC mass to determine the cloud radius (see \S\ref{MOD_dense}), whereas the diffuse clouds have sizes given by the smoothing length of the parent fluid element. \\
   
\noindent {\it Density profile: }
The mass and radius is combined to derive the density profile of each GMC (eq.\,\ref{eq:nH}), while the diffuse gas clouds are assigned uniform densities (\S\ref{MOD_dif}). \\

\noindent {\it FUV radiation field: }
The GMCs experience the combination of the local FUV radiation field flux (eq.\,\ref{FUV}) and the constant background field (eq.\ \ref{G0-diffuse}). The diffuse clouds are only exposed to the background field. In both GMCs and diffuse clouds, we adopt a spectral shape and luminosity corresponding to that of the local Solar neighborhood, which is then scaled by $G_{\rm 0, GMC}$ and $G_{\rm 0, bg}$ respectively.\\

\begin{deluxetable}{lr}
\tablecaption{Adopted parameter values for the GMC grid and diffuse gas \cloudy grids.
\label{table:2}}
\tablewidth{0pt}
\tablehead{
\colhead{Parameter} & \colhead{Grid point values}
}
\startdata
GMCs: & \\
$\log (\mgmc/\msun)$     &	[4.10, 4.28, 4.46, 4.64, 4.82, 5.00] \\	
$\log (G_{\rm 0, GMC}/\FUVMW)$	& 	[0.80, 1.54,  2.28, 3.02, 3.76, 4.50]	\\
$\log ( \Z/\Zsun)$		& 	[-2.00, -1.64, -1.28, -0.92, -0.56, -0.20]	\\
$\log (\Pext/\Pu)$		& 	[5.5, 6.4, 7.3, 8.2, 9.1, 10.0] \\
\hline
Diffuse gas: & \\
$\log (\nH/\cmpc)$       &	[-3.50, -2.76, -2.02, -1.28, -0.54, 0.20] \\
$\log (\R/{\rm kpc})$           	&	[-1.00, -0.76, -0.52, -0.28, -0.04, 0.20] \\
$\log (\Z/\Zsun)$		& 	[-3.00, -2.46, -1.92, -1.38, -0.84, -0.30] \\
$\log (\Tk/{\rm K})$				& 	[2.50, 3.16, 3.82, 4.48, 5.14, 5.80] \\
$G_{\rm 0, bg}/G_{\rm 0,MW}$ & [5, 35]\\
\enddata
\end{deluxetable}

\noindent {\it Element abundances: }
In the \mufasa~simulations, the abundances of He, C, N, O, Ne, Mg, Si, S, Ca and Fe (relative to H) 
are each tracked separately and 
self-consistently calculated at every time-step for each fluid element.
While \cloudy can take an abundance set as input parameter, it is unfeasible to run \cloudy models for every
distinct abundance pattern present in the \mufasa~simulations.
Our default approach is to ascribe the local ISM abundances to a cloud (dense or diffuse), but scale 
all element abundances by the metallicity inherited from the parent fluid element. 
Specifically, we scale the abundances from \cloudy such that 
a sum over the mass fractions of the elements He, C, N, 
O, Ne, Mg, Si, S, Ca and Fe for a fluid
element with solar metallicity gives 0.0134 as expected \citep{asplund09}.\\

\noindent {\it Gas kinetic temperature: }
In the GMCs, \cloudy is allowed to iterate for a temperature. For the diffuse clouds, we keep the temperature fixed at the grid parameter values listed in Table\,\ref{table:2}.

\bigskip

The dust content of each cloud is set to scale linearly with its metallicity, 
normalizing it to the grain abundance in the local ISM. 
The latter is stored in \cloudy as 
graphite and silicate grain abundances 
divided into ten size bins, and corresponds to a 
dust-to-metal (DTM) ratio of 
$0.50$ at solar metallicity. 
Furthermore, dust sublimation is included in the diffuse 
gas models to allow for dust destruction in regions of 
high temperature. 
\cloudy also allows for the inclusion of turbulence by specifying a 
microturbulent velocity. 
For the diffuse clouds we exclude 
turbulence by setting it to 0\,\kms, after a test showed that setting the turbulence to 10\,\kms 
only decreases the \cii luminosities by on average 0.07\,dex. 
For the GMCs we use the velocity dispersion 
obtainable from the cloud radius and pressure \citep[see][]{olsen15}.

\bigskip

At $z\sim6$ the CMB temperature is substantial ($\sim19\,$K), and is included in the calculations of \cloudy with the `CMB command'. The line intensities are corrected to give net line flux above the background continuum, including the diminution effect that happens when upper population levels are sustained by the CMB and reduce the rate of spontaneous de-excitations relative to when excitation happens by collisions only (see \citealt{ferland17} for a more complete description of the CMB treatment in \cloudy version 17). 

\bigskip

\cloudy divides each cloud into a number of shells when solving 
the radiative transfer, heat transfer and chemistry. 
We set the number of shells to $\sim50$, leading to a radial 
resolution in the range $\sim0.01-0.52\,{\rm pc}$ for the GMCs and 
$\sim1-100\,{\rm pc}$ for the diffuse gas clouds.
Due to computational limitations that make it infeasible 
to run \cloudy on every single cloud in our simulations, 
we construct grids of \cloudy models, one grid for GMCs and 
another for the diffuse clouds, and interpolate over those.  
The GMC grid parameters consist of 
$[\mgmc, G_{\rm 0,GMC}, \Z, \Pext]$ 
and the diffuse cloud parameters are 
$[\nH, R, \Z, \Tk]$. 
We ran diffuse gas model grids with $G_{\rm 0, bg}/G_{\rm 0,MW} = 5$, and $35$ (eq.\,\ref{G0-diffuse}), which were the two values most of
the galaxies clustered around. One galaxy (G21) has a diffuse FUV field of $120$ times $G_{\rm 0,MW}$, but adding a further four diffuse gas model grids at $G_{\rm 0}/G_{\rm 0,MW} = 15, 25, 45$, and $120$ to better sample the distribution did not significantly change the line emission properties of the simulations. 
The grid parameter values defining our grids are listed in 
Table \ref{table:2}, and indicated as vertical dashed lines in Figs.~\ref{GMCgridpoints} and \ref{difgridpoints}.
Each grid consists 
of a total of 1296 \cloudy models.  The \cloudy models were 
run on either the Stampede supercomputer at the 
Texas Advanced Computing 
Center\footnote{\url{https://www.tacc.utexas.edu/systems/stampede}}, the Saguaro cluster at 
ASU\footnote{\url{https://researchcomputing.asu.edu/}} 
or the Pleiades supercomputer at 
NASA\footnote{\url{https://www.nas.nasa.gov/hecc/resources/pleiades.html}}.

\begin{figure*}[htb]
\epsscale{1.23}
\hspace*{-0.35cm} \plotone{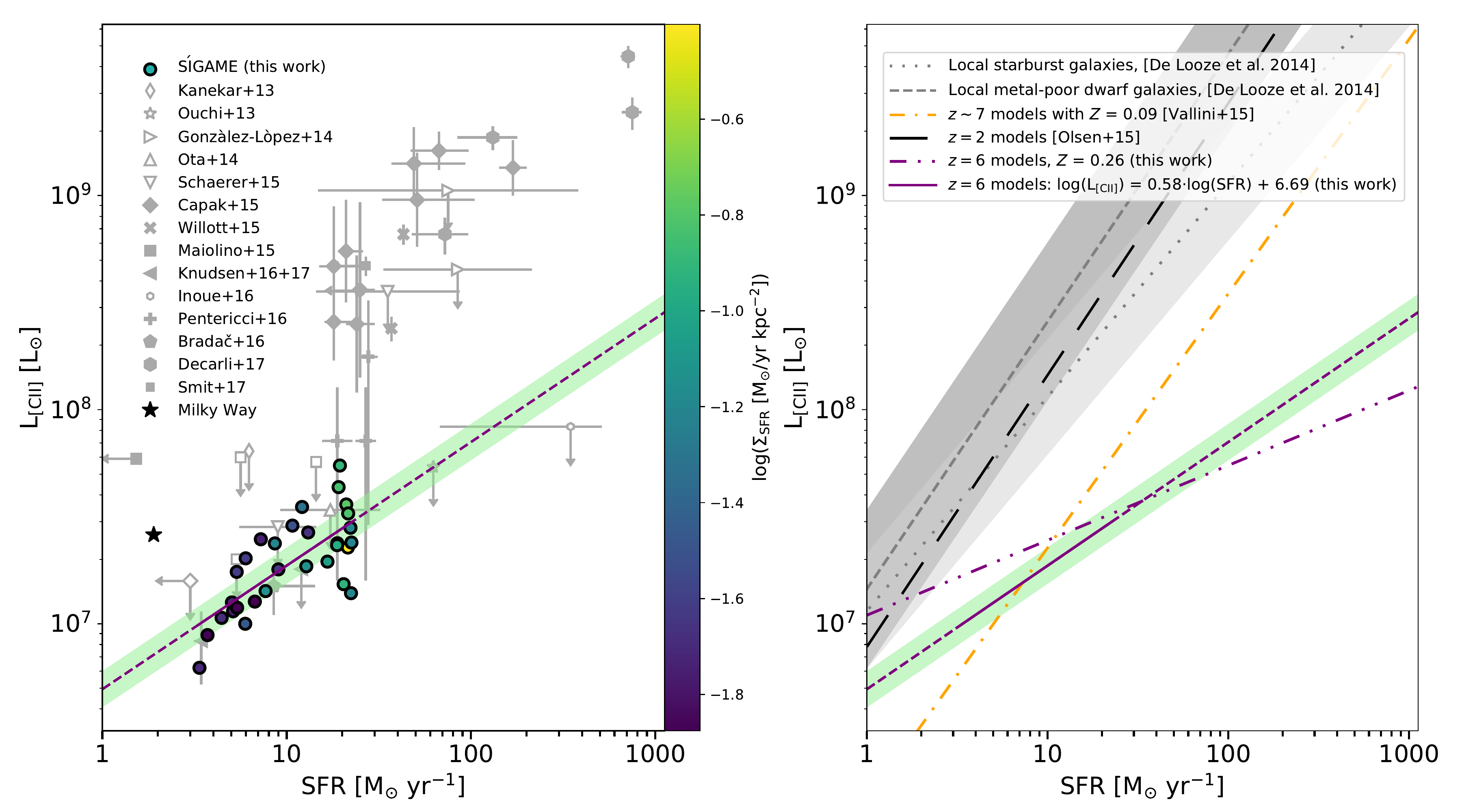}
\caption{{\bf Left}: the \cii luminosity versus SFR for our simulations 
and the green shaded region delineates the scatter ($0.13\,{\rm dex}$) around
this relation. For comparison all \cii observations to date of LBGs and/or LAEs at $z\gs 5$ are shown
(grey symbols, where filled and open symbols indicate detections and 3-$\sigma$ upper limits, respectively). For further details on the observations see \citet{kanekar13,ouchi13,gonzalez-lopez14,ota14,schaerer15,capak15,willott15,maiolino15,inoue16,pentericci16,knudsen16,knudsen17,bradac17,decarli17,smit17}.
For the sake of clarity, we have not included MS0451-H ($z=6.703$) in the plot. This source
has $\Lcii < 3\times 10^5\,\lsun$ and ${\rm SFR \sim 0.4\,\msun\,yr^{-1}}$ \citep{knudsen17}, and lies
about $1\,{\rm dex}$ below our fitted relation. 
All SFRs have been converted to a Chabrier IMF where 
applicable. 
{\bf Right}: a comparison with other, observed and simulated, ${\rm \cii-SFR}$ relations. Shown are the \cii-SFR relations derived by \cite{delooze14} for 
local metal-poor dwarf galaxies (grey dashed line) and 
local starburst galaxies (grey dotted line); 
grey dashed and dotted lines, respectively, with the shaded regions indicating the r.m.s.~scatter of these relations). Also shown are simulated relations of $z=2$ main sequence galaxies (black long dashed line; \citet{olsen15}) and $z\sim7$ galaxies (orange dash-dotted line; \citet{vallini15}).
}
\label{fig:cii_sfr}
\end{figure*}

\section{Simulation results}\label{results}

\subsection{The $L_{\rm [CII]}-$SFR relation at $z\simeq 6$}\label{cii_sfr}
Fig.\,\ref{fig:cii_sfr} (left panel) shows \Lcii against SFR for our 
model galaxies (filled circles). 
There seems to be a correlation between these two quantities, although our simulations only span about a decade in \cii luminosity and SFR and there is significant scatter in \Lcii, 
especially at the high-SFR end.
A Pearson correlation coefficient test yields an $R$-value of
0.64 and a $p$-value of 0.00014 for the likelihood that the observed correlation could arise
by chance if the quantities were uncorrelated.
We can therefore meaningfully fit a log-linear relationship to the simulation points in Fig.\,\ref{fig:cii_sfr}. This yields:
\begin{center}
\scalebox{0.95}{\parbox{\linewidth}{
\begin{eqnarray}
	\log(\Lcii[&&\lsun]) =	(\intercept\pm\interceptdev) \nonumber\\
    &&+(\slope\pm\slopedev)\times\log(\mathrm{SFR}[\sfru]), \label{eq:CII-SFR}
\end{eqnarray}
}}
\end{center}
which is shown as the purple dashed line in Fig.\,\ref{fig:cii_sfr}. 
The uncertainties on the slope and intercept were derived from bootstrapping the 
model results 5000 times.
The r.m.s.\ scatter of the simulated galaxies around this relation is $0.15\,{\rm dex}$ (green shaded region). 

Also shown in Fig.\,\ref{fig:cii_sfr} are the more than two dozen $z\gs 5$ galaxies observed in \cii to date: 23 detections
(filled symbols), and 14 non-detections (upper limits, open symbols). Only normal star forming galaxies (e.g., LBGs and LAEs) have been included in this tally (see Table \ref{table:observations}) -- possibly with the exception of a clump of gas that may not be star forming but was detected in \cii in the vicinity of LBG BDF-3299 \citep{maiolino15}. Obvious AGN-dominated sources and QSOs have been omitted.

\begin{deluxetable*}{l|lcccc}[h!]
\tablecaption{All \cii observations in the literature to date of $z \gs 5$ star forming galaxies. Obvious AGN-dominated sources and QSOs have not been included. Star formation rates and \cii luminosities have been corrected for gravitational magnification ($\mu$). Upper limits are 3-$\sigma$ limits.\label{table:observations}}
\tablewidth{0pt}
\tablehead{\colhead{Name} & \colhead{$z$} & \colhead{SFR\tablenotemark{a} [\sfru]} & \colhead{\Lcii [\lsun]} & \colhead{$\mu$\tablenotemark{b}} & \colhead{Reference} 
}
\startdata
HZ8 			& 5.1533 	& 18\tablenotemark{e} 		& 2.57\e{9}    & ...	& \cite{capak15} \\
HZ7 			& 5.2532 	& 21\tablenotemark{e} 		& 5.50\e{9}    & ...	& \cite{capak15} \\
HZ6 			& 5.2928 	& 49\tablenotemark{e} 		& 1.41\e{10}   & ...	& \cite{capak15} \\
HZ5 			& 5.3089 	& $<3$\tablenotemark{e} 	& $<1.58\e{7}$ & ...	& \cite{capak15} \\
HZ9 			& 5.5410 	& 67\tablenotemark{e} 		& 1.62\e{10}   & ...	& \cite{capak15} \\
HZ3 			& 5.5416 	& 18\tablenotemark{d} 		& 4.68\e{8}    & ...	& \cite{capak15} \\
HZ4 			& 5.5440 	& 51\tablenotemark{e}		& 9.55\e{8}    & ...	& \cite{capak15} \\
HZ10 			& 5.6566 	& 169\tablenotemark{e} 		& 1.34\e{10}   & ...	& \cite{capak15} \\
HZ2 			& 5.6697 	& 25\tablenotemark{e} 		& 3.63\e{8}    & ...	& \cite{capak15} \\
HZ1 			& 5.6885 	& 24\tablenotemark{e} 		& 2.51\e{8}    & ...	& \cite{capak15} \\
A383-5.1		& 6.0274 	& 3.2--3.7\tablenotemark{d} & $<8.3\e{6}$ 		& $11.4\pm1.9$  & \cite{knudsen16} \\
SDSS J0842+1218 comp & 6.0656 	& 131\tablenotemark{f} 	& 1.87\e{9} 	& ...  & \cite{decarli17} \\
WMH5			& 6.0695 	& 43\tablenotemark{c} 		& 6.6\e{8} 	   & 1.27	& \cite{willott15} \\
CFHQ J2100-1715 comp & 6.0796 	& 750\tablenotemark{f} 	& 2.45\e{9} 	& ...  & \cite{decarli17} \\
CLM1 			& 6.1657 	& 37\tablenotemark{c} 		& 2.4\e{8} 	   & ...	& \cite{willott15} \\
PSO J308-21 comp & 6.2485 	& 72\tablenotemark{f} 	& 6.6\e{8} 	& ...  & \cite{decarli17} \\
SDF J132415.7	& 6.541		& 34--211.2\tablenotemark{d} & $<4.52\e{8}$ 	& $<1.1$ & \cite{gonzalez-lopez14} \\	
SDF J132408.3	& 6.554		& 15--375.9\tablenotemark{d} 	& $<10.56\e{8}$   & $<1.1$	& \cite{gonzalez-lopez14} \\
HCM 6A 			& 6.56 		& 6.25\tablenotemark{c} 		& $<0.64\e{8}$ 	& $\approx4.5$  & \cite{kanekar13} \\
PSO J231-20 comp & 6.5900 	& 713\tablenotemark{f} 	& 4.47\e{9} 	& ...  & \cite{decarli17} \\
Himiko			& 6.595		& 62.5\tablenotemark{c} 		& $<0.54\e{8}$ 	& ... & \cite{ouchi13} \\
UDS16291 		& 6.6381 	& 15.8--22.4\tablenotemark{d} 		& 7.15\e{7} 	& ... & \cite{pentericci16} \\
NTTDF6345		& 6.701 	& 25--30.7\tablenotemark{d} 		& 1.77\e{8} 	& ... & \cite{pentericci16} \\
MS0451-H		& 6.703 	& 0.4--0.47\tablenotemark{d} 		& $<3\e{5}$ 	& $100\pm20$ & \cite{knudsen16} \\
RXJ1347:1216 	& 6.7655 	& 8.5\tablenotemark{c} 		& 1.5\e{7} 	& $5\pm0.3$	& \cite{bradac17} \\
A1703-zD1 		& 6.8 		& 5.6--14.3\tablenotemark{d} 	& $<2.83\e{7}$ & $\sim9$	& \cite{schaerer15} \\
COS-2987030247 & 6.8076 	& 16--38.7\tablenotemark{d} 	& 3.6\e{8} 	& ...  & \cite{smit17} \\
SDF-46975 		& 6.844 	& 14.4\tablenotemark{e} 	& $<5.7\e{7}$ 	& ...	& \cite{maiolino15} \\
COS-3018555981 & 6.8540 	& 19--38.2\tablenotemark{d} 	& 4.7\e{8} 	& ...  & \cite{smit17} \\
IOK-1			& 6.96 		& 9.4--31.8\tablenotemark{d}  	& $<3.4\e{7}$ 	& ...	& \cite{ota14} \\	
BDF-512 		& 7.008 	& 5.6\tablenotemark{e} 	& $<6\e{7}$ & ...	&  \cite{maiolino15} \\
BDF-3299 (clump A) 		& 7.107 	& $<1.5$\tablenotemark{e}	& $5.9\e{7}$ & ...	& \cite{maiolino15} \\
BDF-3299 		& 7.109 	& 5.3\tablenotemark{e}	& $<2\e{7}$ & ...	& \cite{maiolino15} \\
COSMOS13679		& 7.1453 	& 23.9--30\tablenotemark{d} 		& 7.12\e{7}   & ...	& \cite{pentericci16} \\
SXDF-NB1006-2 	& 7.2120 	& 347\tablenotemark{c} 	& $<8.4\e{7}$ & ...  & \cite{inoue16} \\
z8-GND-5296 	& 7.508 	& 14.6--85.3\tablenotemark{d} 	& $<3.56\e{8}$ 	& $\sim9$ & \cite{schaerer15} \\
A1689-zD1  		& 7.6031 	& 12\tablenotemark{e} 		& 1.8\e{7} 	& 9.5  & \cite{knudsen17} \\
\enddata
\tablenotetext{a}{\footnotesize SFRs are for a Chabrier IMF. SFRs based on a Kroupa or Salpeter IMF were  corrected by a factor $1.5/1.6$ and $1/1.6$, respectively.}
\tablenotetext{b}{\footnotesize Where the magnification factor is unknown, we assume a value of 1.}
\tablenotetext{c}{\footnotesize SED-based SFR.}
\tablenotetext{d}{\footnotesize Low SFRs are derived from the UV; high SFRs are from the UV combined with an upper limit in the IR.}
\tablenotetext{e}{\footnotesize UV-based SFR.}
\tablenotetext{f}{\footnotesize IR-based SFR.}
\end{deluxetable*}

Based on these data, it would seem there is not a single $L_{\rm [CII]}-{\rm SFR}$ relation. There is a relation defined mainly by the $z\sim 5-6$ LBGs with ${\rm SFR} \gs 20\,\sfru$ observed by \citet{capak15}, \citet{willott15}, \citet{decarli17} and \citet{smit17}, which matches the locally observed ${\rm \Lcii-SFR}$ relation for starburst galaxies \citep{delooze14} and is in agreement with simulations of main-sequence galaxies at $z\sim 2$ \citep{olsen15} (see right-hand panel of Fig.\,\ref{fig:cii_sfr}). Offset from these points, there is a sequence of galaxies with significantly lower \cii luminosities. These are predominantly LAEs with ${\rm SFR < 30\,\msun\,yr^{-1}}$, except for two notable high-SFR sources \citep{ouchi13,inoue16}. 
Since the majority of these sources have in fact only upper limit constraints on their \cii luminosity \citep{kanekar13,ota14,maiolino15,knudsen17} it is not clear whether or not they form a ${\rm \Lcii-SFR}$ relation of their own. 
However, the fact that our simulated galaxies coincide with these \cii-faint sources in the panel suggests that our models are representative of the observations. 
Furthermore, eq.\ \ref{eq:CII-SFR} is consistent with the two high-SFR, \cii upper limits by \citet{ouchi13} and \citet{inoue16}. On the other hand, MS0451-H \citep[$z=6.703$;][]{knudsen17} with its extremely low \cii luminosity ($ < 3\times 10^5\,\lsun$) and SFR (${\rm \sim 0.4\,\msun\,yr^{-1}}$) falls below
eq.\ \ref{eq:CII-SFR} by about $1\,{\rm dex}$ (not shown due to the axis range of the Fig.\,\ref{fig:cii_sfr}). Clearly, more sensitive observations, and of a larger sample of galaxies, are required to delineate a ${\rm \cii-SFR}$ relation at this epoch.

In Fig.\,\ref{fig:cii_sfr} (right panel) we compare our eq.\ \ref{eq:CII-SFR} with the $z\simeq 7$ \cii-SFR relation by \citet{vallini15}. The latter is derived from a fiducial simulation of a single $z\sim 7$ galaxy, and scaling its \Lcii and SFR according to $\Lcii \propto M_{\rm H_2}$ and $\Sigma_{\rm SFR} \propto \Sigma_{\rm H_2}$. Also, the gas phase metallicity of the galaxy is fixed to a constant value throughout the galaxy and  
bears therefore no relation to its star formation history. 
By adopting a set of fixed metallicities, 
\citet{vallini15} arrives at an expression for \Lcii as a function of 
SFR and metallicity.
The orange dash-dotted line in the right-hand side panel shows their 
\cii-SFR relation for a uniform metallicity equal to the 
mean mass-weighted metallicity of
our simulated galaxies ($0.09\times Z_{\rm \odot}$). 
This relation matches our simulated galaxies surprisingly
well given the significant differences between their simulations and ours (see
\S\ref{dis}). 
Clearly, at SFR values outside the range of our simulations
there is significant discrepancy between the \cii-SFR relation by \citet{vallini15}  and ours (eq.\ \ref{eq:CII-SFR}), owing to the much shallower slope of the latter ($\slope\pm\slopedev$ versus $1.2$). 
Extrapolating eq.\ 4 to SFRs well outside the range of the simulations is obviously fraught with danger and
any comparison must be made with caution. This is especially true given that eq.\ 4 is a fit to a set of simulations which span no more than a decade in SFR and $\Lcii$.

\begin{figure*}[t]
\epsscale{1.2}
\plotone{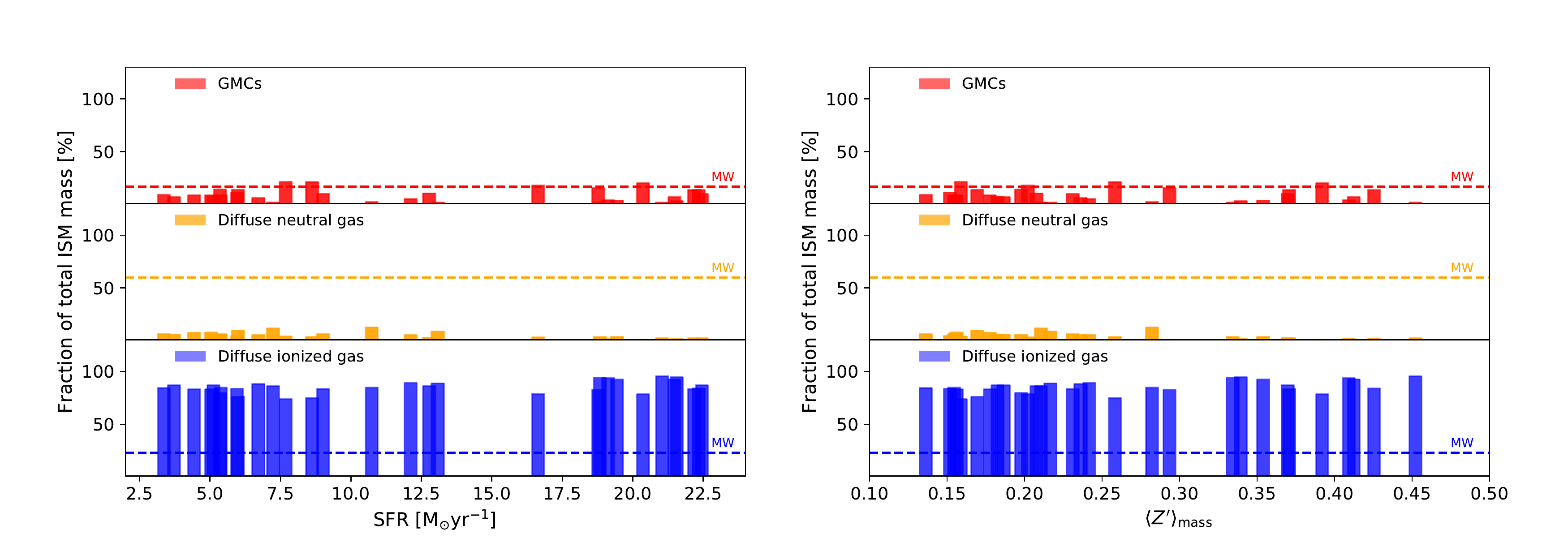}\\
\epsscale{1.2}
\plotone{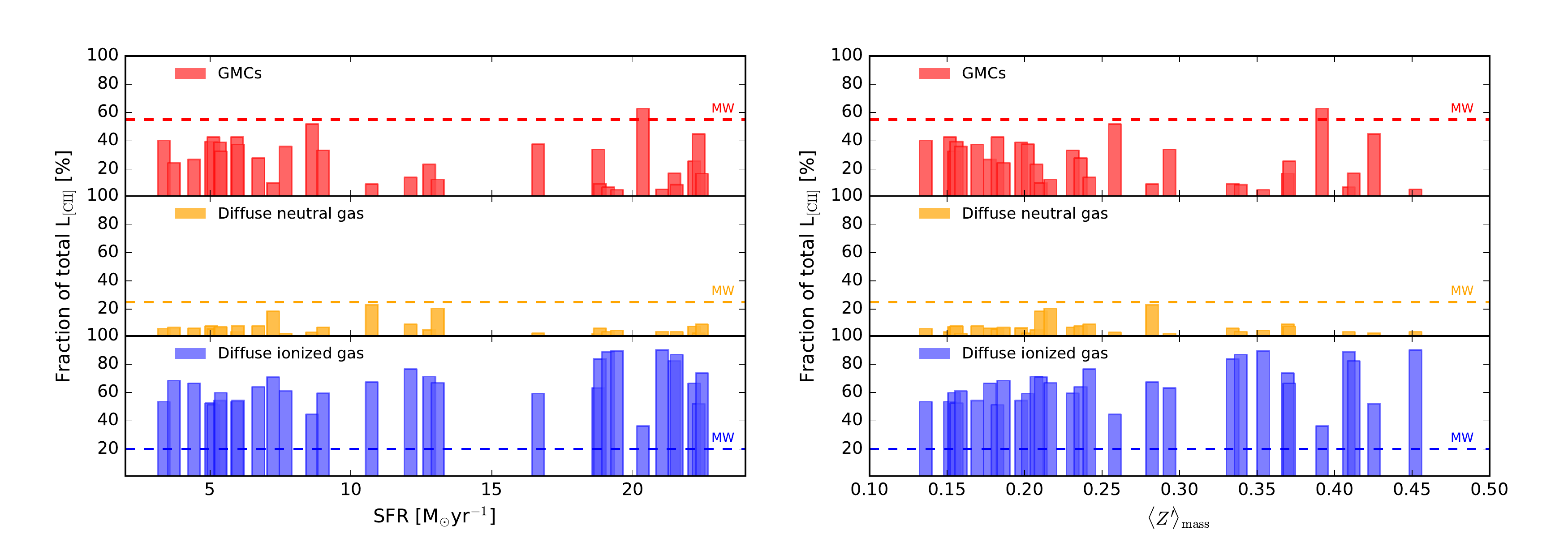}\\
\epsscale{1.2}
\plotone{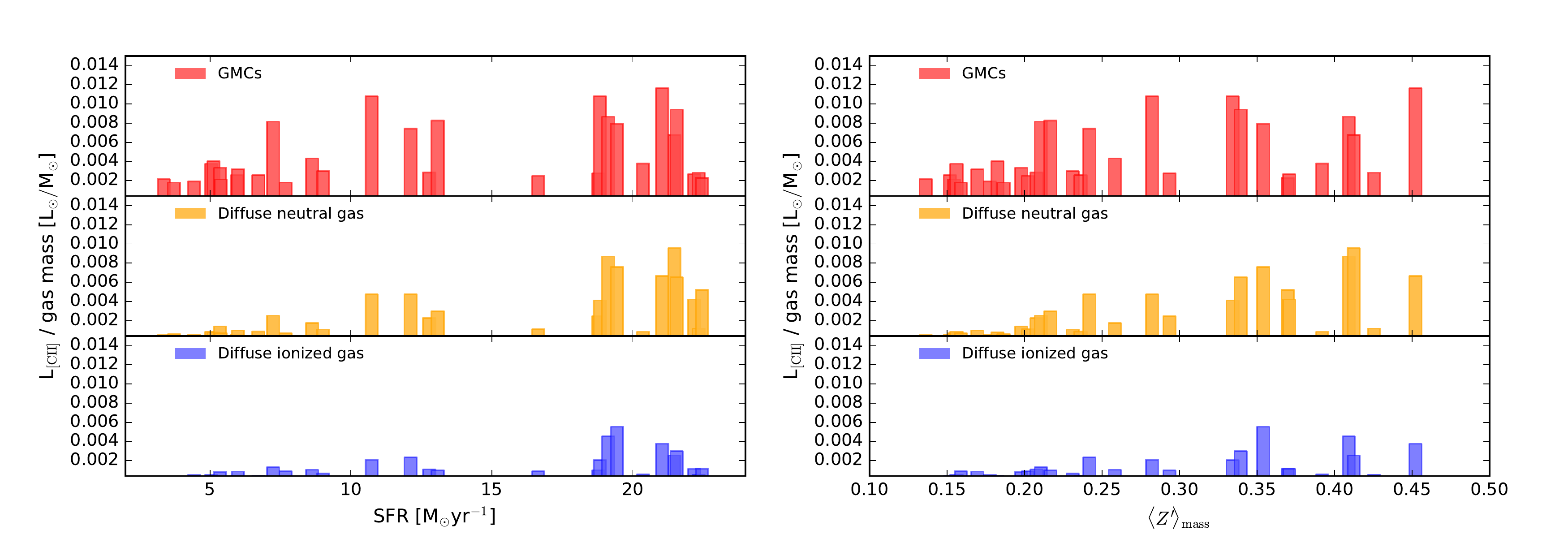}
\caption{{\bf Top}: Mass fractions of the three ISM phases of our simulated galaxies as a function of their total SFRs (left panel) and mass-weighted average metallicities, \Zmw (right panel). The three ISM phases, GMCs, diffuse neutral gas and diffuse ionized gas, are shown as red, orange and blue shaded regions, respectively. The corresponding mass fractions in the MW have been shown with horizontal dashed lines: $17\,\%$ \h2 (red), $60\,\%$ \hi (orange) and $23\,\%$ \hii (blue) within 20\,kpc with numbers from \cite{draine11}.
{\bf Middle:} The contributions from the ISM phases to the total \cii luminosities vs. SFRs (left panel) and \Zmw (right panel). The horizontal dashed lines indicate the corresponding fractions for the MW \citep{pineda14}.
{\bf Bottom}: The \cii luminosity per gas mass of each ISM phase vs. SFRs (left panel) and \Zmw (right panel).
\label{fig:cii_phases}}
\end{figure*}

\subsection{ISM phases and [C\,\textsc{\it II}]}\label{cii_ism} 
The mass fractions of each ISM phase stay relatively constant 
across the 30 simulated galaxies: the GMCs, diffuse neutral 
and diffuse ionized gas 
take up on average $\sim10\,\%$, $\sim6\,\%$ and $\sim85\,\%$ 
of the total ISM mass, respectively, and there is no apparant
trend with SFR or \Zmw (Fig.\ \ref{fig:cii_phases}, top panels).
In contrast, 
the contribution of these ISM phases
to the total \cii luminosity can vary significantly from one
simulated galaxy to another.
The GMCs take up $\sim 5-63\%$ (average $\sim 27\%$), 
the diffuse neutral gas $\sim 1-23\%$ (average $\sim7\%$),
and the diffuse ionized gas $\sim 36-90\%$ (average $\sim 66\%$) 
of the total \cii luminosity. 
The dominant source of \cii in 
the simulated galaxies is either the diffuse ionized gas or the GMCs, with the diffuse neutral gas making up a minor contribution ($\ls 23\%$). 

In Fig.\,\ref{fig:cii_phases} (middle panels) the ISM phase contributions to \Lcii are shown as a function of galaxy SFR (left panel) and \Zmw (right panel). The contribution from the diffuse ionized gas, which is the dominant phase in nearly all the galaxies, tends to increase slightly towards the high-end of the SFR and \Zmw distributions. For the GMCs and the diffuse neutral gas the tendency is in the opposite direction, i.e., their contributions tend to increase toward lower SFR and \Zmw.

\bigskip

In the \citet{vallini15} models with SFRs similar
to our simulations, $\ls 10\%$ of the total \cii emission is coming from diffuse neutral gas, with the remainder coming from PDRs associated with molecular gas. Thus, to the extent that the diffuse neutral gas in the two simulations map onto each other, there is broad agreement about the contribution of this phase to the total \cii luminosity. The lower \cii contributions from molecular gas in our simulations compared to the \citet{vallini15} models ($\sim 37\%$ vs.\ $\gs 90\%$) must, in part at least, be due to the fact that \citet{vallini15} do not include a hot diffuse ionized phase. In contrast, the diffuse ionized gas mass fraction in our simulations is $\sim 85\%$ on average and is responsible for more than half of the total \cii luminosity in most of our simulations. In our Milky Way, dense PDRs and CO-dark H$_2$ gas is responsible for about $\sim 55\%$ of the total \cii emission, while the diffuse ionized gas and diffuse neutral gas contribute $\sim 20\%$ and $25\%$, respectively \citep{pineda14}. 
\cite{cormier12} used \cloudy to model the ISM of the nearby ($z\sim 0.021$) starburst galaxy Haro\,11, which has a metallicity ($\sim0.3\,\Zsun$) and a SFR ($\sim 22\,{\rm \msun\,yr^{-1}}$) matching that of our simulations, and found that $10\,\%$ of the \cii luminosity comes from dense PDRs and about $50\,\%$ from the diffuse ionized medium.

\bigskip

We have seen that the galaxy-to-galaxy variations in the relative \cii contributions from the ISM phases are not matched by their (nearly constant) mass fractions. 
This suggests that the ability of a given ISM phase to shine 
in \cii can change significantly among different galaxies. 
In order to investigate this, we plot in Fig.\,\ref{fig:cii_phases} (bottom panels)
the \cii efficiency, defined as \cii luminosity over gas mass, 
in order of increasing SFR (left panel) and \Zmw (right panel).

Although there are notable exceptions, the \cii efficiencies of all three phases 
tend to increase with SFR and with \Zmw.
The trend with metallicity is easily understood, as
an increase in the carbon abundance will result in a higher abundance of C$^+$. 
The trend with SFR is a combination of the fact that our simulations with higher SFR have higher metallicities (see Table 1) and the fact that, on average, they have stronger FUV radiation fields capable of ionising a larger fraction of the neutral carbon in the GMC and diffuse neutral phase.

\subsection{[OI] and [OIII] at $z\simeq 6$}
\begin{figure}[h]
\epsscale{1.2}
\hspace*{-0.4cm}\plotone{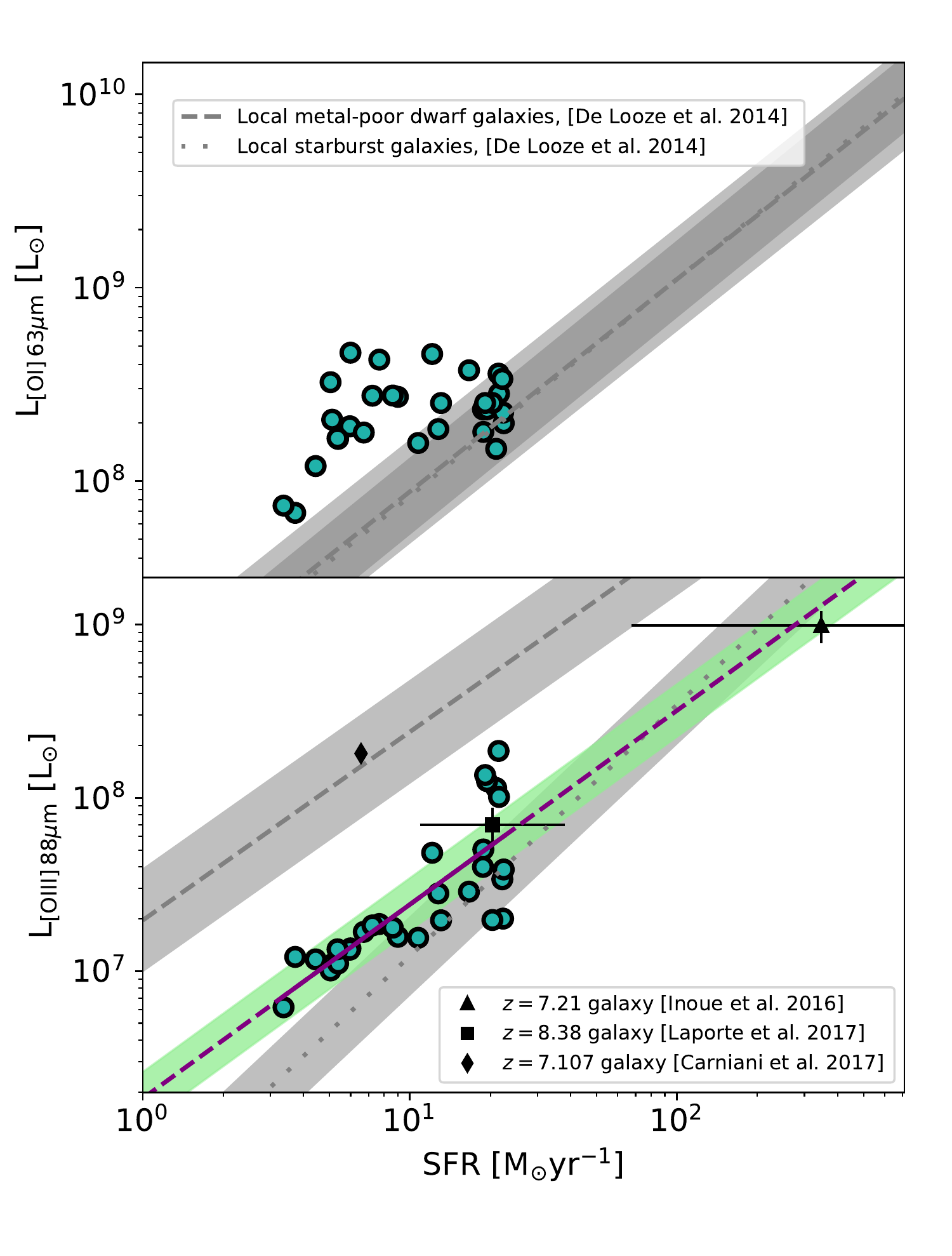}
\caption{
$L_{\rm [OI]}$ vs.\ SFR (top) and $L_{\rm [OIII]}$ vs.\ SFR (bottom) for our simulated galaxies (filled circles) compared to $z \gs 5$ observations of the two lines. 
A log-linear fit to the \oiii simulations is shown by the purple dashed line with a 1$\,\sigma$ scatter ($0.15\,{\rm dex}$) in $L_{\rm [OIII]}$ as indicated by the green shaded region.
Local relations for metal-poor dwarf galaxies (grey dashed lines) and 
local starburst galaxies (grey dotted lines) are shown for comparison \citep{delooze14}. The grey-shaded regions indicate the $\pm 1$-$\sigma$ scatter 
around these relations. Also shown are the detections to date of [OIII] in star forming galaxies at $z\gs 5$ \citep{inoue16,laporte17,carniani17}.
\label{fig:oi}}
\end{figure}

In this section we shall examine the \oi$63\,{\rm \mu m}$ and \oiii$88\,{\rm \mu m}$ line emission from our  simulations. 
Fig.\,\ref{fig:oi} shows \oi and \oiii luminosities 
vs.\ SFRs for our simulated galaxies, along with local $L_{\rm [OI]}-{\rm SFR}$ and $L_{\rm [OIII]}-{\rm SFR}$ relations established for samples of local
metal-poor dwarfs and starburst galaxies \citep{delooze14}. 
Both the \oi and \oiii simulations overlap with the local relations, although the agreement is significantly better for \oiii than for \oi.
In fact, the \oi simulations exhibit no correlation with SFR: a Pearson correlation test of $\log(L_{\rm [OI]})$ vs.\ $\log({\rm SFR})$ gives $R$- and $p$-values of $-0.08$ and $0.69$, respectively. In contrast,
\oiii shows a strong correlation with SFR, and a Pearson correlation test of $\log(L_{\rm [OIII]})$ vs.\ $\log({\rm SFR})$ yields $R=0.79$ and $p=2.5\times 10^{-7}$, respectively. Thus, the \oiii luminosities of our simulations correlate more strongly with SFR than \cii. A log-linear fit yields:
\begin{center}
\scalebox{0.95}{\parbox{\linewidth}{
\begin{eqnarray}
	\log(L_{\rm [OIII]}[&&\lsun]) =	(6.27\pm 0.14)\nonumber\\
    &&+ (1.12\pm 0.17)\times \log(\mathrm{SFR}[\sfru]), \label{eq:OIII-SFR}
\end{eqnarray}
}}
\end{center}
which is shown as the purple dashed line in Fig.\ \ref{fig:oi}. The uncertainties on the slope and intercept were inferred from bootstrapping in a similar way as for \cii (\S\ref{cii_sfr}).
The \oiii detection by \citet{laporte17} of a $z=8.38$ star forming galaxy 
(${\rm SFR} \simeq 20\,{\rm \msun\,yr^{-1}}$) matches extremely well our
simulations with similar SFRs. 
Furthermore, extrapolating 
eq.\,\ref{eq:OIII-SFR} to the SFR ($\sim 250\,{\rm \msun\,yr^{-1}}$)
of the detected LAE at $z=7.21$ \citep[][]{inoue16} 
also results in an excellent match.  
\cite{carniani17} also reported three `blind' \oiii detections in the vicinity of BDF$-$3299 (and at the same redshift). 
In Fig.\,\ref{fig:oi}, we only show the \oiii detection spatially closest to the clump detected in \cii emission \citep{maiolino15}. 
Based on a comparison between models and this \oiii detection, \cite{carniani17} find that an in-situ SFR of $\sim7$\,\sfru is required ($\sim6.6$\,\sfru when converted from a Kroupa to a Chabrier IMF), bringing this object into agreement with the local metal-poor galaxies, but still about 1\,dex above our models.
The \oiii luminosities of the remaining two sources are $2.2$ and $5.8\e{8}$\,\lsun \citep{carniani17}. 

\subsection{[C{\sc ii}] luminosity and global galaxy properties}\label{cii_rel}
In addition to SFR there are likely a number of other galaxy properties affecting the \cii emission. 
For example, in \S\ref{cii_ism} we found that the relative ISM phase contributions to the total \cii luminosity of our simulated galaxies varied not only with SFR but also with average mass-weighted metallicity of the galaxies. This would suggest that the metallicity might have an effect on the $\Lcii-{\rm SFR}$ relation. The simulations by \citet{vallini15} exhibited an increase in the \cii emission with metallicity, which at a basic level is expected since the \cii cooling function scales linearly with the gas phase metallicity \citep{rollig06}. 
Observations of nearby galaxies have also found that surface density of SFR, \SFRsd, can be important for the total amount 
of \cii emitted. For example, \cite{smith17} found that the 
ratio of \cii-to-IR luminosity decreases 
with increasing \SFRsd across six orders of magnitude in \SFRsd. 
Finally, we might expect that the more massive the gas reservoir of a galaxy, the more single ionised carbon is available and, therefore, the brighter the galaxy will shine in \cii. 

\begin{figure*}[t]
\epsscale{1.2}
\plotone{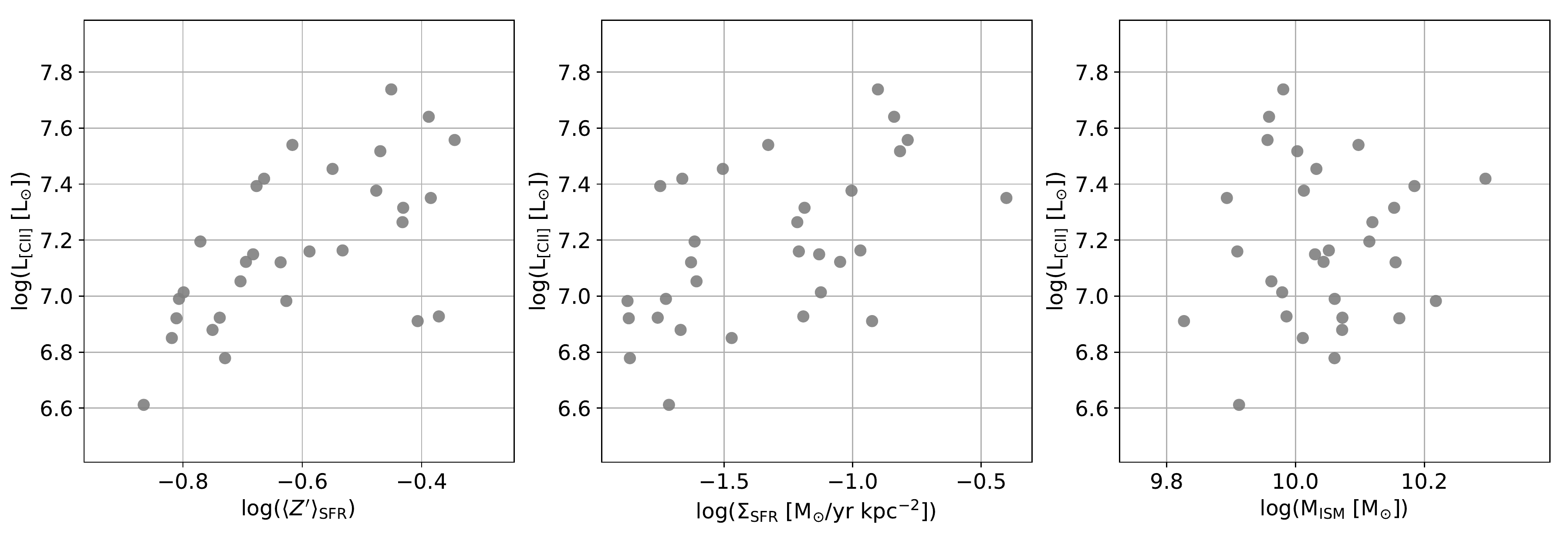}
\caption{\Lcii versus \Zsfr (left), \SFRsd (middle) and \Mism (right) for simulated 
galaxies.
\label{3parameters}}
\end{figure*}

In Fig.\,\ref{3parameters} we have plotted \Lcii of our simulated galaxies against their \Zsfr, \SFRsd 
and \Mism (these quantities are given in Table\,\ref{table:1}, and $\Mism = M_{\rm gas}$). \Lcii appears to correlate with \Zsfr and \SFRsd, whereas there seems to be no discernable correlation with \Mism\, -- perhaps due to the small range spanned in gas mass by our simulations. 

\begin{figure*}[t]
\epsscale{1.2}
\plotone{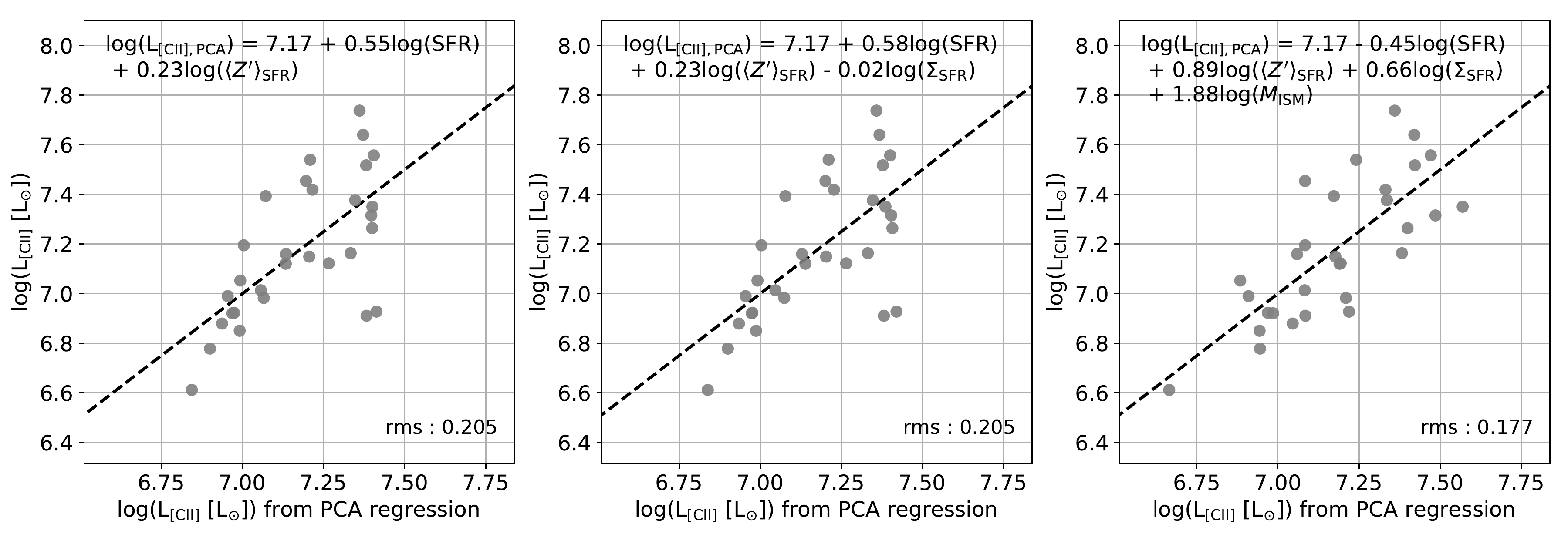}
\caption{Results of a PCA regression to the \cii luminosities of our simulations using different sets of free parameters. {\bf Left panel: }
\SFR and \Zsfr (see eq.\ \ref{eq:pca_regression-3}). {\bf Middle panel:} \SFR, \Zsfr and \SFRsd (see eq.\ \ref{eq:pca_regression-2}). {\bf Right panel:} \SFR, \Zsfr, \SFRsd and \Mism (see eq.\ \ref{eq:pca_regression}). The dashed lines indicate the 1:1 line. The r.m.s.\ scatter of the simulations with respect to eqs.\ \ref{eq:pca_regression}, \ref{eq:pca_regression-2}, and \ref{eq:pca_regression-3} are indicated.}
\label{fig:pca1}
\end{figure*}

In order to better examine the dependence of \Lcii on the above quantities, and to account for their inter-dependance, we  perform a principle component analysis 
\citep[PCA;][]{jolliffe12} analysis. 
Normalizing the logarithm of the aforementioned quantities (and SFR) to zero mean leaves the following 
variables for the PCA:
\begin{eqnarray}
	&&x_1 = \log(\SFR) - 1.02 \nonumber \\
	&&x_2 = \log(\Zsfr) - (-0.61) \nonumber \\
	&&x_3 = \log(\SFRsd) - (-1.33) \nonumber \\ 
	&&x_4 = \log(\Mism) - 10.04 \label{eq:pca_par} 
\end{eqnarray}
The resulting principle components in this four dimensional 
parameter space are:
\begin{eqnarray}
&&\mathrm{PC}_1 = 0.52x_1 + 0.28x_2 + 0.80x_3 - 0.10x_4 \nonumber \\
&&\mathrm{PC}_2 = 0.64x_1 + 0.36x_2 - 0.48x_3 + 0.47x_4 \nonumber \\
&&\mathrm{PC}_3 = 0.53x_1 - 0.45x_2 - 0.27x_3 - 0.66x_4 \nonumber \\
&&\mathrm{PC}_4 = 0.18x_1 - 0.76x_2 + 0.22x_3 + 0.57x_4. \label{eq:pc1} 
\end{eqnarray}
88\,\% of the sample variance is contained within 
the eigenvector PC$_1$ which is dominated by and increases 
strongly with SFR and \SFRsd. 
PC$_2$, PC$_3$ and PC$_4$ contain $\sim9$, $0.6$ 
and $2.3\,\%$ of the variance, respectively. 
Thus of the global galaxy properties considered, the star formation activity is the most important factor for driving the \cii luminosity.

Keeping all four principle components, a regression 
can be made to:
\begin{eqnarray}
\log(\Lcii) = &&\beta_0 + \beta_1 \mathrm{PC}_1 + 
\beta_2 \mathrm{PC}_2 \nonumber\\
&&+ \beta_3 \mathrm{PC}_3 + \beta_4 \mathrm{PC}_4,
\end{eqnarray}
leading to the following relation for \Lcii:
\begin{eqnarray}
\log(\Lcii) &&= 7.17 - 0.45 \log(\SFR) + 0.89 \log(\Zsfr) \nonumber \\
&&+ 0.66 \log(\SFRsd) + 1.88 \log(\Mism).
\label{eq:pca_regression}
\end{eqnarray}
In Fig.\,\ref{fig:pca1} (right hand panel) we show the \cii luminosities obtained by applying eq.\ \ref{eq:pca_regression} to our model galaxies plotted against their true \cii luminosities from the simulations. The scatter between the true \cii luminosities of the galaxies and eq.\ \ref{eq:pca_regression} is only $0.18\,{\rm dex}$, suggesting that
eq.\ \ref{eq:pca_regression} captures most of the \cii dependencies on global galaxy parameters.
Repeating the above PCA and regression analysis but using only (\SFR, \Zsfr, \SFRsd)
and (\SFR, \Zsfr) as free parameter sets, we obtain:
\begin{eqnarray}
\log(\Lcii) &&= 7.17 - 0.58 \log(\SFR) + 0.23 \log(\Zsfr) \nonumber \\
&&- 0.02 \log(\SFRsd),
\label{eq:pca_regression-2}
\end{eqnarray}
and
\begin{eqnarray}
\log(\Lcii) &&= 7.17 + 0.55 \log(\SFR) + 0.23 \log(\Zsfr), \nonumber \\
\label{eq:pca_regression-3}
\end{eqnarray}
respectively. It is seen that limiting the free parameters to (${\rm SFR}$, \Zsfr,\SFRsd) (middle panel) or (${\rm SFR}$,\Zsfr) (left panel), deteriorates the correlations somewhat, with a slight increase in the scatter as a result ($0.21\,{\rm dex}$).

\section{Discussion}\label{dis}
\subsection{Simulation robustness}
In this section we examine the effects on the simulation outcomes when changing some of the assumptions build into \sigame (\S\ref{MOD}). To this end, we ran \sigame on the same 30 \mufasa~simulations presented in \S\ref{SPH} but with changes made to: 1) the dust-to-metals mass ratio, 2) the slope of the GMC mass spectrum, or 3) the element abundances.

\bigskip

\noindent{\it The dust-to-metals mass ratio.} The simulations presented in this paper have adopted a DTM 
mass ratio of $0.5$, close to $\sim0.62$ of the MW  
\citep[the latter derived from an average over 
243 lines of sight through the 
local part of our Galaxy;][]{wiseman17}.
However, recent studies at high redshifts have shown that 
dust production is less efficient at low metallicities resulting 
in DTM mass ratios significantly below that of the MW 
in low metallicity systems \citep{vladilo04,decia13,decia16,wiseman17}.

\begin{figure*}[t]
\epsscale{1.3}
\hspace*{-1.09cm}\plotone{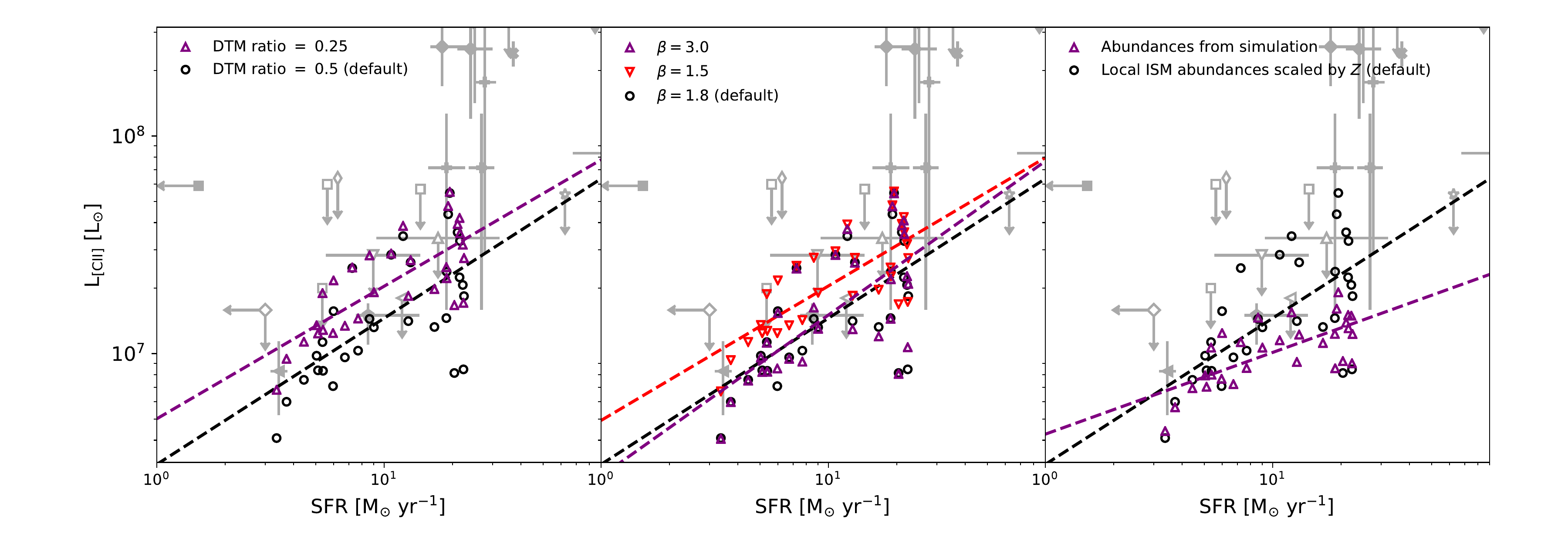}
\caption{The result of deviating from the default assumptions of \sigame. 
 Purple triangles show the simulated galaxies in the \Lcii-SFR diagram after: reducing the DTM ratio by a factor 2 (left); changing the GMC mass spectrum to one that is more bottom- 
or top-heavy (middle); adopting abundances that better match the cosmological simulation (right). Black circles show the location of model galaxies with the default assumptions 
of \sigame. Observations are shown with grey symbols as in Fig.\,\ref{fig:cii_sfr}. 
\label{fig:alt_CII_sfr}}
\end{figure*}

We ran \sigame using new grids of \cloudy models with a DTM ratio of $0.25$, i.e., half of that adopted in \S\ref{MOD_clo}.
The effect is to increase the \cii luminosities of the simulated galaxies by {\it $\sim 43\%$ on average or 0.15\,dex and the change to the \Lcii-SFR
relation is therefore significant
(Fig.\,\ref{fig:alt_CII_sfr}, left  panel).} The 
increase in \Lcii is primarily coming from
the GMCs, where decreasing the 
amount of dust reduces the shielding of 
the gas from FUV radiation, thereby allowing for larger \cplus envelopes. 

\bigskip

\noindent{\it The GMC mass distribution.}
In high-$z$ environments the GMCs are likely to
differ from the Galactic mass spectrum (slope $\beta = 1.8$),
which was adopted for our simulations (\S\ref{MOD_dense}). 
In our Local Group alone, where GMC masses can be measured, significant variations in $\beta$ have been found: from $\beta = 1.5$ in the inner MW to $\beta = 2.1$ and $\beta = 2.9$ in the outer MW and in M33, respectively \citep{rosolowsky05,blitz07}.
Running \sigame with a `bottom-heavy' ($\beta = 3.0$) GMC mass spectrum has practically no impact on the \cii emission of the galaxies, but switching to a `top-heavy' ($\beta = 1.5$) GMC mass spectrum increases \Lcii by an average of $44\,\%$ or $0.15$\,dex (Fig.\,\ref{fig:alt_CII_sfr}, middle panel), similar to the case of reduced DTM ratio.

\bigskip

\begin{figure*}[b!]
  \centering
  \begin{turn}{-90}
  \hspace{-2cm}
  \begin{minipage}[b][][b]{22cm}
  \includegraphics[width=22cm]{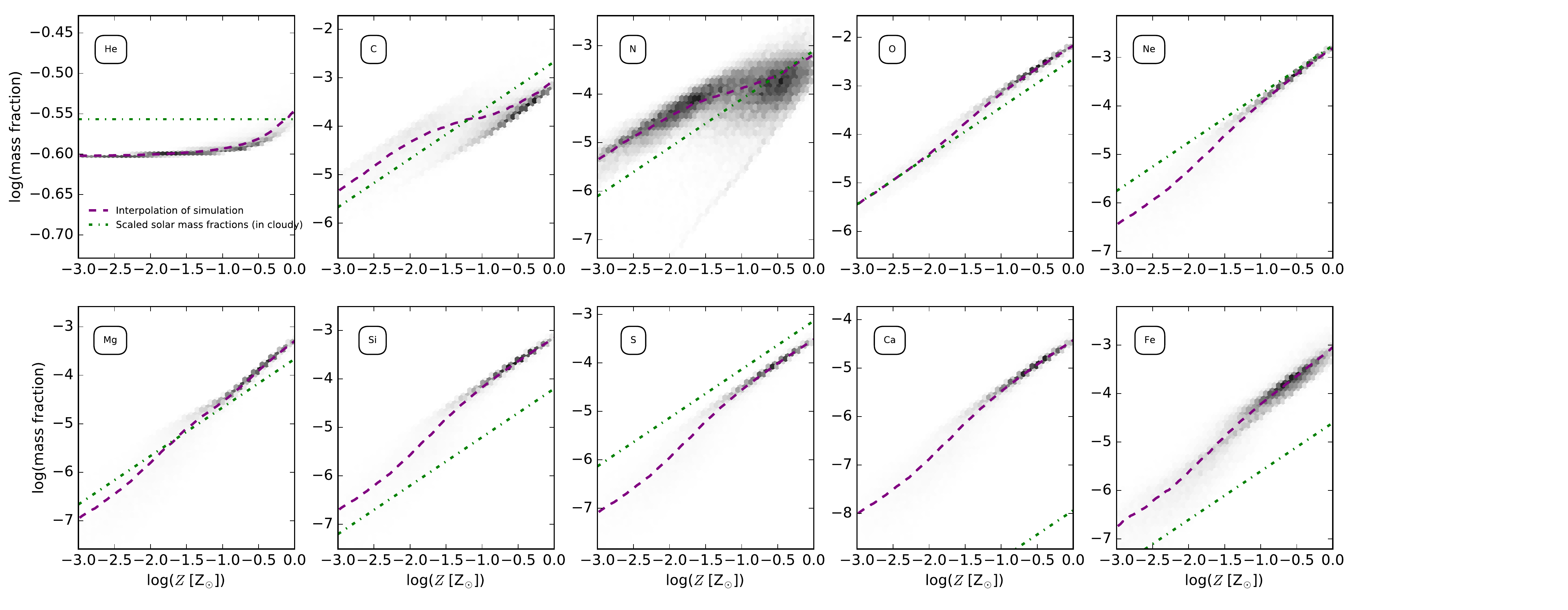}
    \caption{Grey contours represent the mass fractions 
      of the elements tracked in the \mufasa~simulation 
      set as function of the metallicity 
      of individual fluid elements for a subsample of 
      three randomly selected model galaxies used in this 
      work. Using the local ISM abundances 
      stored in \cloudy and scaled by metallicity gives 
      the green dash-dotted line. These are our default 
      abundances in \sigame for this project. 
      We also make a grid of cloudy models in which each element 
      shown here is scaled to 
      match the simulation as represented by the purple 
      dashed lines which consist of running means of the values of the gas fluid elements.}
  \end{minipage}
  \label{mass_fraction_scaling}
  \end{turn}
\end{figure*}

\noindent{\it Abundances.} 
As we saw in \S\ref{MOD_clo} \sigame ascribes to a cloud (dense or diffuse) the element abundances of the local ISM, but scaled by the metallicity of the parent fluid element. This ensures a one-to-one relation between the abundance
of an element and the metallicity, which is required in order to
keep the number of \cloudy models at a manageable level. 
These relations (one for each element) can easily be converted to relations between element mass fractions and metallicity and are plotted (green dash-dotted lines) in Fig.\,\ref{mass_fraction_scaling}, which shows element mass fractions against metallicity for the fluid elements in three of the \mufasa~galaxy simulations.
It is seen that this approach does not match the element mass fractions of the simulations particularly well.

As a result, an alternative approach was devised, that would ensure that the element abundances provided as input to \cloudy
did in fact reflect the typical element mass fractions in the \mufasa~simulations. In this approach
a one-to-one relation between the mass fraction of an element and the metallicity is established by simply doing
spline-fits to the average mass fractions within 100
bins in metallicity (purple dashed curve in Fig.\,\ref{mass_fraction_scaling}). It is then straightforward to infer the corresponding element abundances (as a function of metallicity).
Adopting this approach tends to lower the \cii luminosities of our simulations, although in some cases there is actually a very slight increase.
The net outcome is a shallower $\Lcii-{\rm SFR}$ relation (Fig.\,\ref{fig:alt_CII_sfr}, right panel), but one which remains consistent with the \cii-faint observations.
The lower \cii luminosities are due to the fact that over the metallicity range spanning the majority of both dense and diffuse clouds in the simulations ($-1.5 \ls \log Z/Z_{\rm \odot} \ls 0$; Fig.\,\ref{GMCgridpoints} and \ref{difgridpoints}), the default local ISM abundances (green dash-dotted lines) overshoot the average simulation carbon abundances (purple dashed curves).
\\


\subsection{Reconciling [CII]-luminous sources at $z\sim 6$ with our simulations}
In the previous section we saw that our simulation findings are relatively robust against 
changes in some of the model assumptions. In particular, we saw that it is  difficult to
increase the \cii emission from the simulations. 
This then begs the question of how to account for the \cii-luminous LBGs detected by \citet{capak15}, \citet{willott15} and \citet{smit17} (and the luminous QSO companions detected by \citealt{decarli17}), which lie significantly above the $\Lcii-{\rm SFR}$ relation defined by our simulations and the many \cii-faint sources \citep[e.g.,][]{maiolino15,bradac17}.

\smallskip

\noindent{\it Observational uncertainties.}
The SFRs of galaxies at $z\sim 6$ are
derived via either conversion from IR luminosities or 
rest-frame UV continuum luminosities (corrected for dust). 
Although there can be significant uncertainties associated with estimating
SFRs of high-$z$ galaxies, in the case of the \citet{capak15} and \citet{willott15} 
LBGs they would have had to be underestimated by an order of magnitude if this was the reason for the above discrepancy.
\citet{capak15} estimates a systematic uncertainty of $0.3\,{\rm dex}$ on the
IR luminosities, which are inferred from only a single long-wavelength ($\lambda_{\rm obs} = 0.85-1\,{\rm mm}$) data-point and scaling a modified black body laws to it with dust temperatures in the range $25-45\,{\rm K}$, spectral indices of $1.2-2.0$, and Wien power-law slopes of $1.5-2.5$. 

We have made an estimate of the dust temperatures in our model galaxies using the dust radiative transfer package 
\powderday \ \citep{narayanan15a,narayanan17b}
to generate dust emission SEDs for the central halos that 
we extract galaxies from. 
\powderday builds off of {\sc hyperion} \citep{robitaille11a,robitaille12a}, {\sc fsps} \citep{conroy09b,conroy10a}, and {\sc yt} \citep{turk11a}. In short, \powderday \ generates stellar spectra for all the stars formed in the simulation using their ages and metallicites, and computing their SEDs as simple stellar populations using {\sc fsps}. The metal properties of the model galaxies are projected onto an adaptive grid with an octree memory structure using {\sc yt}, and then the stellar SEDs are propagated through the dusty ISM using {\sc hyperion} as the dust radiative transfer solver. This process is iterated upon in a Monte Carlo fashion until the radiation field and dust temperatures have achieved equilibrium. 
Since we only extract central 
galaxies, these SEDs will be dominated by the light from our simulated galaxies.
The resulting dust temperatures lie in the range $50-72$\,K, i.e., significantly warmer than the \citet{capak15} sources. If $>50\,{\rm K}$ is a more appropriate temperature range for the dust in $z\sim 6$ LBGs, it would imply that their IR luminosities had been underestimated. Whether this also implies higher star formation rates is less clear. We note that this is a good bit warmer than the typical dust temperature of simulations of comparable luminosity galaxies at $z\sim 0-2$ \citep[e.g.][]{narayanan10a,narayanan11}; this is due principally to the low dust contents and hard radiation fields in low-metallicity galaxies at $z > 5$.

\smallskip

\noindent{\it AGN.}
The possibility that the \cii-brightness of the \citet{capak15} and \citet{willott13} sources could be due to
the presence of AGN seems unlikely since normal LBGs at $z\sim 6$ are expected to have moderate
mass black holes ($\ls 10^8\,{\rm \msun}$) that would not dominate the overall energetics. Also, \cii observations
of QSOs at $z\sim 6$ (and at lower redshifts) tend to show lower, not higher, $L_{\rm [CII]}/L_{\rm IR}$ values compared to
star forming galaxies \citep{maiolino05,venemans12,wang13,zhao16}.

\smallskip

\noindent{\it Gas mass fractions.} 
Is it possible that our simulations and the \cii-faint sources have significantly
lower gas mass fractions, and thus smaller gas reservoirs that can emit in \cii, than the \cii-bright LBGs?
In \S\ref{cii_ism} we found that the molecular and diffuse ionized gas phases contribute about equally to the total \cii luminosities of simulations with ${\rm SFR}\ls 10\,{\rm \msun\,yr^{-1}}$, while at higher SFRs the ionized gas tends to dominate. 
Taking these findings at face value would suggest that the \cii-bright LBGs may have higher fractions of either ionized or molecular gas, or both, than our simulations and \cii-faint sources.

\begin{figure}[t]
\epsscale{1.2}
\hspace*{-0.15cm}\plotone{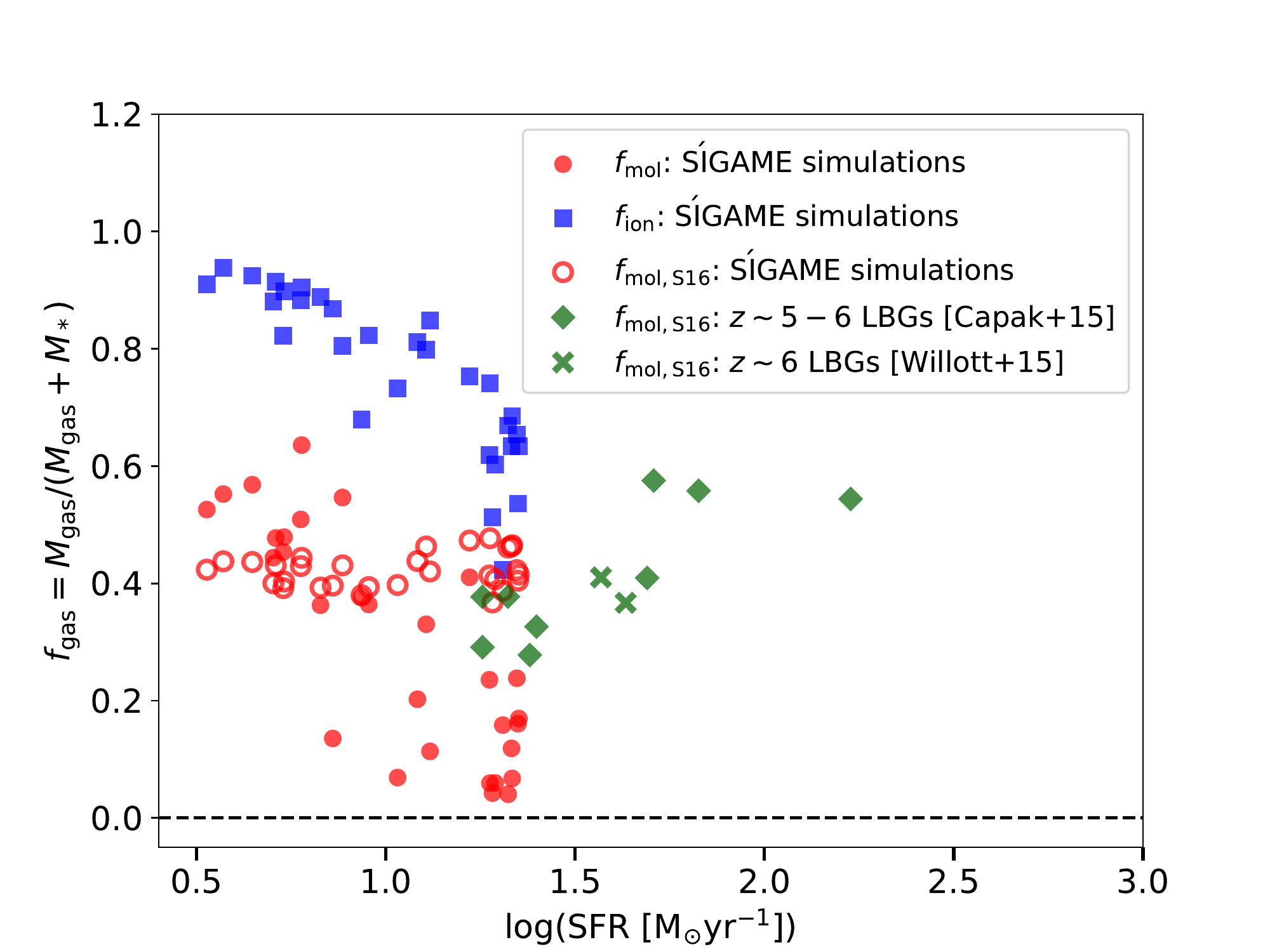}
\caption{Molecular gas mass fractions of our simulations and of the $z\sim5-6$ LBGs
observed by \cite{capak15} and \cite{willott15}. The filled and open red circles indicate true fractions inferred directly from the simulations and fractions inferred from
the parametrisations in \cite{scoville16}, respectively. Applying this parametrisation to the 
$z\sim5-6$ LBGs gives the molecular gas mass fractions shown by the
green diamonds and crosses. Also shown are the true ionized gas mass fraction of our simulations (blue filled squares).
\label{fig:f_mol}}
\end{figure}

To investigate this we first show in Fig.\,\ref{fig:f_mol} the molecular gas mass fractions ($f_{\rm mol}$) of our simulated galaxies as calculated directly from Table 1.
The molecular gas mass fractions decrease 
with SFR, going from $\sim 0.6$ at the lowest SFR to $\sim 0.1$ at the highest SFRs (red filled circles in Fig.\,\ref{fig:f_mol}). For the \citet{capak15} and \cite{willott15} LBGs we do not have direct estimates of their gas mass fractions. Instead we apply to them the parametrisations provided by \citet{scoville16}, which give $f_{\rm mol}$ as a function of $z$, $M_{\rm *}$, and SFR (and denominated $f_{\rm mol,S16}$ in the figure). The resulting fractions for the LBGs, shown as green diamonds and crosses in Fig.\,\ref{fig:f_mol}, are significantly higher ($f_{\rm mol,S16}\sim 0.4-0.6$) than those of our simulations at ${\rm SFR \gs 10\,\msun\,yr^{-1}}$ ($f_{\rm mol}\sim 0.1$). 
Applying the same parametrisations to our simulated galaxies yields $f_{\rm mol,S16} \sim 0.4$, which matches well with the \citet{capak15} and \cite{willott15} sources. 
This would mean roughly $4\times$ higher $f_{\rm mol}$ at the highest SFRs of our models, which translates into a $6\times$ higher molecular gas mass for a fixed stellar mass. 
For comparison we also show the ionized gas mass fractions, $f_{\rm ion}$, of our model galaxies, which at present cannot be compared to the $z\sim5-6$ LBGs of \cite{capak15} and \cite{willott15}, but we note a decrease in $f_{\rm ion}$ with SFR similar to that of $f_{\rm mol}$. 

In adopting the relation from \cite{scoville16}, we are making three major assumptions: 1) that the relation created from galaxies out to $z = 5.89$ can be extrapolated to $z\sim6$, 2) that \Mstar and SFR are derived in similar fashions from observations as from our simulations, and 3) that the actual molecular gas masses in our simulations correspond to the dust-derived molecular ISM masses used in \cite{scoville16}. However, if all these assumptions are valid, then it implies that the range of $f_{\rm mol}$ in our model galaxies does not reach the higher $f_{\rm mol}$ values derived for observed galaxies galaxies close to $z=6$. This will tend to make our models underpredict the \cii emission since we find that the \cii efficiency of molecular regions is generally higher than diffuse regions of the ISM (cf. Fig\,\ref{fig:cii_phases}).

\smallskip

\noindent{\it Metallicities.}
In \S\ref{cii_rel} we saw that simulations with higher average metallicities
also tend to have higher \cii luminosities (Fig.\,\ref{3parameters} and eq.\,\ref{eq:pca_regression}). Our simulations all have metallicities well below solar ($\Zsfr \ls 0.45$) -- and less than half the metallicity of the local metal-poor dwarf galaxies studied by \citet{delooze14} --
which may
therefore partly account for their \cii faintness and the discrepancy with the \citet{capak15} sources which are likely to have higher metallicities.
 
\begin{figure}[t]
\epsscale{1.35}
\hspace*{-0.5cm}\plotone{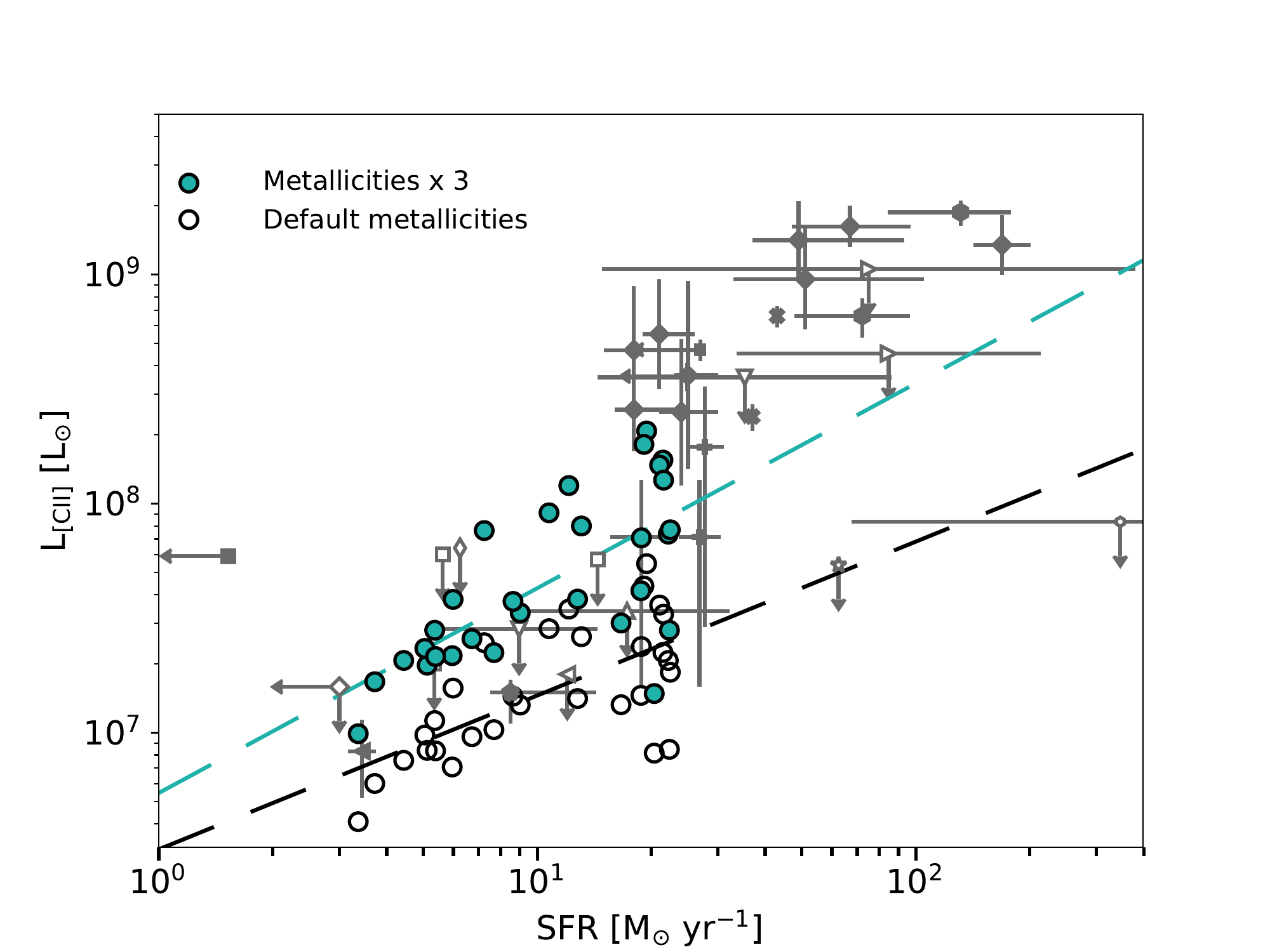}
\caption{The effect of boosting the (default) metallicities of each fluid element
in our simulations by a factor of three results in galaxy \cii luminosities (filled cyan circles)
that are on average $\sim 2.5-3\times$ higher.  Even so, this does not bring the simulations fully into accord
with the \cii-bright detections at $z\simeq 6$, although there is an overlap with observations for some simulations. The dashed cyan line shows the log-linear fit to the simulations with high metallicities.
\label{fig:cii_sfr_Z}}
\end{figure}

Fig.\,\ref{fig:cii_sfr_Z} shows the \cii-SFR relation obtained when scaling the metallicities of our simulations by a factor of three leading to SFR-weighted metallicities of $0.4$ to $1.4$ solar and 
a mass-metallicity relation for our model galaxies close to that observed at 
$z\sim1.3-2.3$ for galaxies of similar masses \citep{erb06,henry13,sanders15}. 
This is therefore an extreme case, in which our $z\sim6$ galaxies have 
already achieved typical metallicities at $z\sim1.3-2.3$. 
In order to test this case, new \cloudy grids were calculated 
for both the dense and diffuse gas. 
The net effect of raising the metallicities 
is to increase the \cii luminosities of our simulations by $0.4$\,dex on average, 
thereby bringing them better in line with the \cii detections of \citet{capak15}
at ${\rm SFR} \sim 20\,{\rm \msun\,yr^{-1}}$. 
This picture is consistent with the 
relatively low Ly$\alpha$ 
line strength measured for the $z\sim5-6$ detections 
\citep{capak15,willott15,pentericci16}, since the Ly$\alpha$ strength decreases with the dust amount and hence metallicity \citep[e.g.][]{pirzkal07,yang17}, except in cases where the ISM is very inhomogeneous \citep[e.g.][]{giavalisco96}, though see also the model results by \cite{laursen13}.

We note that the $\sim 0.4\,{\rm dex}$ increase in \cii luminosity of our simulations is mainly coming
from the neutral and diffuse ionized gas where an increase in the metallicity
results in a higher abundance of \cplus. For the GMCs, an increase in the metallicity
results in a higher degree of UV-attenuation by dust and therefore thinner \cplus layers.
This is the same effect we saw in Fig.\,\ref{fig:cii_phases} where the fraction of the total \cii luminosity
coming from GMCs decreases with increasing metallicity of the simulations.

\section{Conclusion}\label{con}
We have applied an updated version of
\sigame to \mufasa~simulations in order to
model the \cii, \oi, and \oiii
line emission from $z\sim6$ galaxies on the main sequence with stellar masses $\sim (0.7-8)\times 10^9\,{\rm \msun}$ and metallicities $\sim (0.1 - 0.4)\times \Zsun$. 

The simulations are able to reproduce observations of \cii-faint star forming galaxies at 
$z\gs 5$ -- i.e., galaxies with \cii luminosities (many of which are upper limits) that are $\sim6-32\times$ lower than expected from local \Lcii-SFR relations and from samples of \cii-bright galaxies at $z\sim5-6$ \citep{capak15,willott15}.
Extrapolating a log-linear fit to our simulations to higher SFRs ($\sim 50-300\,{\rm \msun\,yr^{-1}}$) results in agreement with 
observations of \cii-faint galaxies at these higher SFRs.

The \oi and \oiii luminosities from the simulated galaxies are in broad agreement with the $L_{\rm [OI]}$-SFR and $L_{\rm [OIII]}$-SFR relations of local starburst galaxies \citep{delooze14}, as well as with two of the three existing $z\gs 5$ detections of \oiii to date \citep{inoue16,laporte17}.  

\bigskip

Dividing the \cii luminosities of our simulations into contributions from the different ISM phases, we find that the \cii emission predominantly comes from the diffuse ionized gas and the GMCs, 
which on average contribute by $\sim 66\%$ and $\sim 27\%$, respectively, while
the diffuse neutral phase contributes   by $\sim 7\%$.
In terms of mass, the three phases constitute on average $\sim 85\%$, 
$\sim 10\%$, and $\sim 10\%$ of the total ISM mass in our simulations.
Thus, the GMCs are the most efficient \cii emitters of the ISM phases, suggesting that the molecular gas fraction plays a role in whether a galaxy is \cii-faint or \cii-bright.

A principle component analysis shows that \Lcii primarily depends on \SFRsd and SFR. Furthermore, including metallicity in the set of free global parameters reduces the scatter between \Lcii from the PCA parametrization and \Lcii from the simulations. In our models, the ISM mass is not an important parameter in setting \Lcii.

The modeling presented in this paper suggest that the \cii-faint $z\gtrsim5$ main sequence galaxies, including \cii non-detections, are likely the result of low metallicities
and low molecular gas fractions.
More observations of FIR emission lines at high redshift together with more precise SFR determinations are needed in order to compare better with models such as ours, yet at the same time we have to work towards modeling galaxies with more dynamic range in physical parameters such as metallicity and SFR.

\section*{Acknowledgments}
The authors thank the anonymous referee for a careful reading of the manuscript and insightful comments that greatly improved the quality of this work. The authors thank Sangeeta Malhotra and Kristian Finlator for fruitful discussion. 
The authors acknowledge the Texas Advanced Computing Center (TACC) at The University of Texas at Austin for providing high performance computing resources that have contributed to the research results reported within this paper (\url{http://www.tacc.utexas.edu}). 
In addition, this work benefited from resources provided by the NASA High-End Computing (HEC) Program through the NASA Advanced Supercomputing (NAS) Division, and from computational resources and technical expertise provided by ASU Research Computing, Tempe, Arizona.
The hydrodynamic simulations published here were run on the University of Florida HiPerGator facility, and the authors acknowledge the University of Florida Research Computing for providing computational resources and support that have contributed to the research results reported in this
publication.  
Partial support for DN was provided by NSF AST-1009452, AST-1445357, NASA HST AR-13906.001 from the Space Telescope Science Institute, which is operated by the Association of University for Research in Astronomy, Incorporated, under NASA Contract NAS5-26555, and a Cottrell College Science Award, awarded by the Research Corporation for Science Advancement. 

\bibliographystyle{aasjournal}
\bibliography{bibs}

\begin{thebibliography}{}
\expandafter\ifx\csname natexlab\endcsname\relax\def\natexlab#1{#1}\fi

\bibitem[{{Accurso} {et~al.}(2017){Accurso}, {Saintonge}, {Bisbas}, \&
  {Viti}}]{accurso17}
{Accurso}, G., {Saintonge}, A., {Bisbas}, T.~G., \& {Viti}, S. 2017, \mnras,
  464, 3315

\bibitem[{{Asplund} {et~al.}(2009){Asplund}, {Grevesse}, {Sauval}, \&
  {Scott}}]{asplund09}
{Asplund}, M., {Grevesse}, N., {Sauval}, A.~J., \& {Scott}, P. 2009, \araa, 47,
  481

\bibitem[{{Blitz} {et~al.}(2007){Blitz}, {Fukui}, {Kawamura}, {Leroy},
  {Mizuno}, \& {Rosolowsky}}]{blitz07}
{Blitz}, L., {Fukui}, Y., {Kawamura}, A., {et~al.} 2007, Protostars and Planets
  V, 81

\bibitem[{{Bonato} {et~al.}(2014){Bonato}, {Negrello}, {Cai}, {De Zotti},
  {Bressan}, {Lapi}, {Gruppioni}, {Spinoglio}, \& {Danese}}]{bonato14}
{Bonato}, M., {Negrello}, M., {Cai}, Z.-Y., {et~al.} 2014, \mnras, 438, 2547

\bibitem[{{Brada{\v c}} {et~al.}(2017){Brada{\v c}}, {Garcia-Appadoo}, {Huang},
  {Vallini}, {Quinn Finney}, {Hoag}, {Lemaux}, {Borello Schmidt}, {Treu},
  {Carilli}, {Dijkstra}, {Ferrara}, {Fontana}, {Jones}, {Ryan}, {Wagg}, \&
  {Gonzalez}}]{bradac17}
{Brada{\v c}}, M., {Garcia-Appadoo}, D., {Huang}, K.-H., {et~al.} 2017, \apjl,
  836, L2

\bibitem[{{Brauher} {et~al.}(2008){Brauher}, {Dale}, \& {Helou}}]{brauher08}
{Brauher}, J.~R., {Dale}, D.~A., \& {Helou}, G. 2008, \apjs, 178, 280

\bibitem[{{Bruzual} \& {Charlot}(2003)}]{bruzual03a}
{Bruzual}, G., \& {Charlot}, S. 2003, \mnras, 344, 1000

\bibitem[{{Capak} {et~al.}(2015){Capak}, {Carilli}, {Jones}, {Casey},
  {Riechers}, {Sheth}, {Carollo}, {Ilbert}, {Karim}, {Lefevre}, {Lilly},
  {Scoville}, {Smolcic}, \& {Yan}}]{capak15}
{Capak}, P.~L., {Carilli}, C., {Jones}, G., {et~al.} 2015, \nat, 522, 455

\bibitem[{{Carniani} {et~al.}(2017){Carniani}, {Maiolino}, {Pallottini},
  {Vallini}, {Pentericci}, {Ferrara}, {Castellano}, {Vanzella}, {Grazian},
  {Gallerani}, {Santini}, {Wagg}, \& {Fontana}}]{carniani17}
{Carniani}, S., {Maiolino}, R., {Pallottini}, A., {et~al.} 2017, ArXiv
  e-prints, arXiv:1701.03468

\bibitem[{{Chabrier}(2003)}]{chabrier03}
{Chabrier}, G. 2003, PASP, 115, 763

\bibitem[{{Chomiuk} \& {Povich}(2011)}]{chomiuk11}
{Chomiuk}, L., \& {Povich}, M.~S. 2011, \aj, 142, 197

\bibitem[{{Conroy} {et~al.}(2009){Conroy}, {Gunn}, \& {White}}]{conroy09b}
{Conroy}, C., {Gunn}, J.~E., \& {White}, M. 2009, \apj, 699, 486

\bibitem[{{Conroy} {et~al.}(2010){Conroy}, {White}, \& {Gunn}}]{conroy10a}
{Conroy}, C., {White}, M., \& {Gunn}, J.~E. 2010, \apj, 708, 58

\bibitem[{{Cormier} {et~al.}(2012){Cormier}, {Lebouteiller}, {Madden}, {Abel},
  {Hony}, {Galliano}, {Baes}, {Barlow}, {Cooray}, {De Looze}, {Galametz},
  {Karczewski}, {Parkin}, {R{\'e}my}, {Sauvage}, {Spinoglio}, {Wilson}, \&
  {Wu}}]{cormier12}
{Cormier}, D., {Lebouteiller}, V., {Madden}, S.~C., {et~al.} 2012, \aap, 548,
  A20

\bibitem[{{Dav{\'e}} {et~al.}(2017){Dav{\'e}}, {Rafieferantsoa}, {Thompson}, \&
  {Hopkins}}]{dave17}
{Dav{\'e}}, R., {Rafieferantsoa}, M.~H., {Thompson}, R.~J., \& {Hopkins}, P.~F.
  2017, \mnras, 467, 115

\bibitem[{{Dav{\'e}} {et~al.}(2016){Dav{\'e}}, {Thompson}, \&
  {Hopkins}}]{dave16a}
{Dav{\'e}}, R., {Thompson}, R., \& {Hopkins}, P.~F. 2016, \mnras, 462, 3265

\bibitem[{{De Cia} {et~al.}(2016){De Cia}, {Ledoux}, {Mattsson}, {Petitjean},
  {Srianand}, {Gavignaud}, \& {Jenkins}}]{decia16}
{De Cia}, A., {Ledoux}, C., {Mattsson}, L., {et~al.} 2016, \aap, 596, A97

\bibitem[{{De Cia} {et~al.}(2013){De Cia}, {Ledoux}, {Savaglio}, {Schady}, \&
  {Vreeswijk}}]{decia13}
{De Cia}, A., {Ledoux}, C., {Savaglio}, S., {Schady}, P., \& {Vreeswijk}, P.~M.
  2013, \aap, 560, A88

\bibitem[{{De Looze} {et~al.}(2014){De Looze}, {Cormier}, {Lebouteiller},
  {et~al.}}]{delooze14}
{De Looze}, I., {Cormier}, D., {Lebouteiller}, V., {et~al.} 2014, \aap, 568,
  A62

\bibitem[{{Decarli} {et~al.}(2017){Decarli}, {Walter}, {Venemans},
  {Ba{\~n}ados}, {Bertoldi}, {Carilli}, {Fan}, {Farina}, {Mazzucchelli},
  {Riechers}, {Rix}, {Strauss}, {Wang}, \& {Yang}}]{decarli17}
{Decarli}, R., {Walter}, F., {Venemans}, B.~P., {et~al.} 2017, \nat, 545, 457

\bibitem[{{D{\'{\i}}az-Santos} {et~al.}(2013){D{\'{\i}}az-Santos}, {Armus},
  {Charmandaris}, {et~al.}}]{diaz-santos13}
{D{\'{\i}}az-Santos}, T., {Armus}, L., {Charmandaris}, V., {et~al.} 2013, \apj,
  774, 68

\bibitem[{{Diaz-Santos} {et~al.}(2017){Diaz-Santos}, {Armus}, {Charmandaris},
  {Lu}, {Stierwalt}, {Stacey}, {Malhotra}, {van der Werf}, {Howell}, {Privon},
  {Mazzarella}, {Goldsmith}, {Murphy}, {Barcos-Munoz}, {Linden}, {Inami},
  {Larson}, {Evans}, {Appleton}, {Iwasawa}, {Lord}, {Sanders}, \&
  {Surace}}]{diaz-santos17}
{Diaz-Santos}, T., {Armus}, L., {Charmandaris}, V., {et~al.} 2017, ArXiv
  e-prints, arXiv:1705.04326

\bibitem[{{Draine}(2011)}]{draine11}
{Draine}, B.~T. 2011, {Physics of the Interstellar and Intergalactic Medium}

\bibitem[{{Erb} {et~al.}(2006){Erb}, {Shapley}, {Pettini}, {Steidel}, {Reddy},
  \& {Adelberger}}]{erb06}
{Erb}, D.~K., {Shapley}, A.~E., {Pettini}, M., {et~al.} 2006, \apj, 644, 813

\bibitem[{{Ferkinhoff} {et~al.}(2010){Ferkinhoff}, {Hailey-Dunsheath},
  {Nikola}, {Parshley}, {Stacey}, {Benford}, \& {Staguhn}}]{ferkinhoff10}
{Ferkinhoff}, C., {Hailey-Dunsheath}, S., {Nikola}, T., {et~al.} 2010, \apjl,
  714, L147

\bibitem[{{Ferland} {et~al.}(2013){Ferland}, {Porter}, {van Hoof},
  {et~al.}}]{ferland13}
{Ferland}, G.~J., {Porter}, R.~L., {van Hoof}, P.~A.~M., {et~al.} 2013, RxMAA,
  49, 137

\bibitem[{{Ferland} {et~al.}(2017){Ferland}, {Chatzikos}, {Guzm{\'a}n},
  {Lykins}, {van Hoof}, {Williams}, {Abel}, {Badnell}, {Keenan}, {Porter}, \&
  {Stancil}}]{ferland17}
{Ferland}, G.~J., {Chatzikos}, M., {Guzm{\'a}n}, F., {et~al.} 2017, ArXiv
  e-prints, arXiv:1705.10877

\bibitem[{{Fischer} {et~al.}(1999){Fischer}, {Luhman}, {Satyapal},
  {Greenhouse}, {Stacey}, {Bradford}, {Lord}, {Brauher}, {Unger}, {Clegg},
  {Smith}, {Melnick}, {Colbert}, {Malkan}, {Spinoglio}, {Cox}, {Harvey},
  {Suter}, \& {Strelnitski}}]{fischer99}
{Fischer}, J., {Luhman}, M.~L., {Satyapal}, S., {et~al.} 1999, \apss, 266, 91

\bibitem[{{Giavalisco} {et~al.}(1996){Giavalisco}, {Koratkar}, \&
  {Calzetti}}]{giavalisco96}
{Giavalisco}, M., {Koratkar}, A., \& {Calzetti}, D. 1996, \apj, 466, 831

\bibitem[{{Goldsmith} {et~al.}(2012){Goldsmith}, {Langer}, {Pineda}, \&
  {Velusamy}}]{goldsmith12}
{Goldsmith}, P.~F., {Langer}, W.~D., {Pineda}, J.~L., \& {Velusamy}, T. 2012,
  \apjs, 203, 13

\bibitem[{{Gonz{\'a}lez-L{\'o}pez} {et~al.}(2014){Gonz{\'a}lez-L{\'o}pez},
  {Riechers}, {Decarli}, {Walter}, {Vallini}, {Neri}, {Bertoldi}, {Bolatto},
  {Carilli}, {Cox}, {da Cunha}, {Ferrara}, {Gallerani}, \&
  {Infante}}]{gonzalez-lopez14}
{Gonz{\'a}lez-L{\'o}pez}, J., {Riechers}, D.~A., {Decarli}, R., {et~al.} 2014,
  \apj, 784, 99

\bibitem[{{Gullberg} {et~al.}(2015){Gullberg}, {De Breuck}, {Vieira},
  {et~al.}}]{gullberg15}
{Gullberg}, B., {De Breuck}, C., {Vieira}, J.~D., {et~al.} 2015, \mnras, 449,
  2883

\bibitem[{{Hahn} \& {Abel}(2011)}]{hahn11a}
{Hahn}, O., \& {Abel}, T. 2011, \mnras, 415, 2101

\bibitem[{{Hayward} \& {Smith}(2015)}]{hayward15a}
{Hayward}, C.~C., \& {Smith}, D.~J.~B. 2015, \mnras, 446, 1512

\bibitem[{{Helou} {et~al.}(2001){Helou}, {Malhotra}, {Hollenbach}, {Dale}, \&
  {Contursi}}]{helou01}
{Helou}, G., {Malhotra}, S., {Hollenbach}, D.~J., {Dale}, D.~A., \& {Contursi},
  A. 2001, \apjl, 548, L73

\bibitem[{{Henry} {et~al.}(2013){Henry}, {Scarlata}, {Dom{\'{\i}}nguez},
  {Malkan}, {Martin}, {Siana}, {Atek}, {Bedregal}, {Colbert}, {Rafelski},
  {Ross}, {Teplitz}, {Bunker}, {Dressler}, {Hathi}, {Masters}, {McCarthy}, \&
  {Straughn}}]{henry13}
{Henry}, A., {Scarlata}, C., {Dom{\'{\i}}nguez}, A., {et~al.} 2013, \apjl, 776,
  L27

\bibitem[{{Herrera-Camus} {et~al.}(2015){Herrera-Camus}, {Bolatto}, {Wolfire},
  {et~al.}}]{herrera-camus15}
{Herrera-Camus}, R., {Bolatto}, A.~D., {Wolfire}, M.~G., {et~al.} 2015, \apj,
  800, 1

\bibitem[{{Hopkins}(2015)}]{hopkins15}
{Hopkins}, P.~F. 2015, \mnras, 450, 53

\bibitem[{{Inoue} {et~al.}(2016){Inoue}, {Tamura}, {Matsuo}, {Mawatari},
  {Shimizu}, {Shibuya}, {Ota}, {Yoshida}, {Zackrisson}, {Kashikawa}, {Kohno},
  {Umehata}, {Hatsukade}, {Iye}, {Matsuda}, {Okamoto}, \&
  {Yamaguchi}}]{inoue16}
{Inoue}, A.~K., {Tamura}, Y., {Matsuo}, H., {et~al.} 2016, Science, 352, 1559

\bibitem[{{Iwamoto} {et~al.}(1999){Iwamoto}, {Brachwitz}, {Nomoto},
  {Kishimoto}, {Umeda}, {Hix}, \& {Thielemann}}]{iwamoto99a}
{Iwamoto}, K., {Brachwitz}, F., {Nomoto}, K., {et~al.} 1999, \apjs, 125, 439

\bibitem[{{Jiang} {et~al.}(2016){Jiang}, {Finlator}, {Cohen}, {Egami},
  {Windhorst}, {Fan}, {Dav{\'e}}, {Kashikawa}, {Mechtley}, {Ouchi},
  {Shimasaku}, \& {Cl{\'e}ment}}]{jiang16}
{Jiang}, L., {Finlator}, K., {Cohen}, S.~H., {et~al.} 2016, \apj, 816, 16

\bibitem[{{Jolliffe}(2002)}]{jolliffe12}
{Jolliffe}, I. 2002, Principal component analysis (New York: Springer Verlag)

\bibitem[{{Kanekar} {et~al.}(2013){Kanekar}, {Wagg}, {Ram Chary}, \&
  {Carilli}}]{kanekar13}
{Kanekar}, N., {Wagg}, J., {Ram Chary}, R., \& {Carilli}, C.~L. 2013, \apjl,
  771, L20

\bibitem[{{Kapala} {et~al.}(2015){Kapala}, {Sandstrom}, {Groves},
  {et~al.}}]{Kapala15}
{Kapala}, M.~J., {Sandstrom}, K., {Groves}, B., {et~al.} 2015, \apj, 798, 24

\bibitem[{{Katz} {et~al.}(2017){Katz}, {Kimm}, {Sijacki}, \&
  {Haehnelt}}]{katz17}
{Katz}, H., {Kimm}, T., {Sijacki}, D., \& {Haehnelt}, M.~G. 2017, \mnras, 468,
  4831

\bibitem[{{Kennicutt} \& {Evans}(2012)}]{kennicutt12}
{Kennicutt}, R.~C., \& {Evans}, N.~J. 2012, \araa, 50, 531

\bibitem[{{Knudsen} {et~al.}(2016){Knudsen}, {Richard}, {Kneib}, {Jauzac},
  {Cl{\'e}ment}, {Drouart}, {Egami}, \& {Lindroos}}]{knudsen16}
{Knudsen}, K.~K., {Richard}, J., {Kneib}, J.-P., {et~al.} 2016, \mnras, 462, L6

\bibitem[{{Knudsen} {et~al.}(2017){Knudsen}, {Watson}, {Frayer}, {Christensen},
  {Gallazzi}, {Micha{\l}owski}, {Richard}, \& {Zavala}}]{knudsen17}
{Knudsen}, K.~K., {Watson}, D., {Frayer}, D., {et~al.} 2017, \mnras, 466, 138

\bibitem[{{Krumholz} \& {Gnedin}(2011)}]{krumholz11}
{Krumholz}, M.~R., \& {Gnedin}, N.~Y. 2011, \apj, 729, 36

\bibitem[{{Krumholz} {et~al.}(2009){Krumholz}, {McKee}, \&
  {Tumlinson}}]{krumholz09}
{Krumholz}, M.~R., {McKee}, C.~F., \& {Tumlinson}, J. 2009, \apj, 693, 216

\bibitem[{{Laporte} {et~al.}(2017){Laporte}, {Ellis}, {Boone}, {Bauer},
  {Qu{\'e}nard}, {Roberts-Borsani}, {Pell{\'o}}, {P{\'e}rez-Fournon}, \&
  {Streblyanska}}]{laporte17}
{Laporte}, N., {Ellis}, R.~S., {Boone}, F., {et~al.} 2017, \apjl, 837, L21

\bibitem[{{Laursen} {et~al.}(2013){Laursen}, {Duval}, \&
  {{\"O}stlin}}]{laursen13}
{Laursen}, P., {Duval}, F., \& {{\"O}stlin}, G. 2013, \apj, 766, 124

\bibitem[{{Leitherer} {et~al.}(2014){Leitherer}, {Ekstr{\"o}m}, {Meynet},
  {Schaerer}, {Agienko}, \& {Levesque}}]{leitherer14}
{Leitherer}, C., {Ekstr{\"o}m}, S., {Meynet}, G., {et~al.} 2014, \apjs, 212, 14

\bibitem[{{Luhman} {et~al.}(2003){Luhman}, {Satyapal}, {Fischer}, {Wolfire},
  {Sturm}, {Dudley}, {Lutz}, \& {Genzel}}]{luhman03}
{Luhman}, M.~L., {Satyapal}, S., {Fischer}, J., {et~al.} 2003, \apj, 594, 758

\bibitem[{{Luhman} {et~al.}(1998){Luhman}, {Satyapal}, {Fischer}, {Wolfire},
  {Cox}, {Lord}, {Smith}, {Stacey}, \& {Unger}}]{luhman98}
---. 1998, \apjl, 504, L11

\bibitem[{{Maiolino} {et~al.}(2005){Maiolino}, {Cox}, {Caselli},
  {et~al.}}]{maiolino05}
{Maiolino}, R., {Cox}, P., {Caselli}, P., {et~al.} 2005, \aap, 440, L51

\bibitem[{{Maiolino} {et~al.}(2015){Maiolino}, {Carniani}, {Fontana},
  {Vallini}, {Pentericci}, {Ferrara}, {Vanzella}, {Grazian}, {Gallerani},
  {Castellano}, {Cristiani}, {Brammer}, {Santini}, {Wagg}, \&
  {Williams}}]{maiolino15}
{Maiolino}, R., {Carniani}, S., {Fontana}, A., {et~al.} 2015, \mnras, 452, 54

\bibitem[{{Malhotra} {et~al.}(1997){Malhotra}, {Helou}, {Stacey},
  {et~al.}}]{malhotra97}
{Malhotra}, S., {Helou}, G., {Stacey}, G., {et~al.} 1997, \apjl, 491, L27

\bibitem[{{Malhotra} {et~al.}(2001){Malhotra}, {Kaufman}, {Hollenbach},
  {et~al.}}]{malhotra01}
{Malhotra}, S., {Kaufman}, M.~J., {Hollenbach}, D., {et~al.} 2001, \apj, 561,
  766

\bibitem[{{Mu{\~n}oz} \& {Furlanetto}(2014)}]{munoz14}
{Mu{\~n}oz}, J.~A., \& {Furlanetto}, S.~R. 2014, \mnras, 438, 2483

\bibitem[{{Muratov} {et~al.}(2015){Muratov}, {Kere{\v s}},
  {Faucher-Gigu{\`e}re}, {Hopkins}, {Quataert}, \& {Murray}}]{muratov15a}
{Muratov}, A.~L., {Kere{\v s}}, D., {Faucher-Gigu{\`e}re}, C.-A., {et~al.}
  2015, \mnras, 454, 2691

\bibitem[{{Narayanan} {et~al.}(2017){Narayanan}, {Dave}, {Johnson}, {Thompson},
  {Conroy}, \& {Geach}}]{narayanan17b}
{Narayanan}, D., {Dave}, R., {Johnson}, B., {et~al.} 2017, arXiv/1705.05858,
  arXiv:1705.05858

\bibitem[{{Narayanan} {et~al.}(2011){Narayanan}, {Krumholz}, {Ostriker}, \&
  {Hernquist}}]{narayanan11}
{Narayanan}, D., {Krumholz}, M., {Ostriker}, E.~C., \& {Hernquist}, L. 2011,
  \mnras, 418, 664

\bibitem[{{Narayanan} \& {Krumholz}(2017)}]{narayanan17}
{Narayanan}, D., \& {Krumholz}, M.~R. 2017, \mnras, 467, 50

\bibitem[{{Narayanan} {et~al.}(2010){Narayanan}, {Dey}, {Hayward}, {Cox},
  {Bussmann}, {Brodwin}, {Jonsson}, {Hopkins}, {Groves}, {Younger}, \&
  {Hernquist}}]{narayanan10a}
{Narayanan}, D., {Dey}, A., {Hayward}, C.~C., {et~al.} 2010, \mnras, 407, 1701

\bibitem[{{Narayanan} {et~al.}(2015){Narayanan}, {Turk}, {Feldmann},
  {Robitaille}, {Hopkins}, {Thompson}, {Hayward}, {Ball},
  {Faucher-Gigu{\`e}re}, \& {Kere{\v s}}}]{narayanan15a}
{Narayanan}, D., {Turk}, M., {Feldmann}, R., {et~al.} 2015, \nat, 525, 496

\bibitem[{{Nomoto} {et~al.}(2006){Nomoto}, {Tominaga}, {Umeda}, {Kobayashi}, \&
  {Maeda}}]{nomoto06a}
{Nomoto}, K., {Tominaga}, N., {Umeda}, H., {Kobayashi}, C., \& {Maeda}, K.
  2006, Nuclear Physics A, 777, 424

\bibitem[{{Olsen} {et~al.}(2015){Olsen}, {Greve}, {Narayanan}, {Thompson},
  {Toft}, \& {Brinch}}]{olsen15}
{Olsen}, K.~P., {Greve}, T.~R., {Narayanan}, D., {et~al.} 2015, \apj, 814, 76

\bibitem[{{Oppenheimer} \& {Dav{\'e}}(2008)}]{oppenheimer08}
{Oppenheimer}, B.~D., \& {Dav{\'e}}, R. 2008, \mnras, 387, 577

\bibitem[{{Ota} {et~al.}(2014){Ota}, {Walter}, {Ohta}, {Hatsukade}, {Carilli},
  {da Cunha}, {Gonz{\'a}lez-L{\'o}pez}, {Decarli}, {Hodge}, {Nagai}, {Egami},
  {Jiang}, {Iye}, {Kashikawa}, {Riechers}, {Bertoldi}, {Cox}, {Neri}, \&
  {Weiss}}]{ota14}
{Ota}, K., {Walter}, F., {Ohta}, K., {et~al.} 2014, \apj, 792, 34

\bibitem[{{Ouchi} {et~al.}(2013){Ouchi}, {Ellis}, {Ono}, {et~al.}}]{ouchi13}
{Ouchi}, M., {Ellis}, R., {Ono}, Y., {et~al.} 2013, \apj, 778, 102

\bibitem[{{Pallottini} {et~al.}(2017){Pallottini}, {Ferrara}, {Gallerani},
  {Vallini}, {Maiolino}, \& {Salvadori}}]{pallottini17}
{Pallottini}, A., {Ferrara}, A., {Gallerani}, S., {et~al.} 2017, \mnras, 465,
  2540

\bibitem[{{Papadopoulos} {et~al.}(2011){Papadopoulos}, {Thi}, {Miniati}, \&
  {Viti}}]{papa11}
{Papadopoulos}, P.~P., {Thi}, W.-F., {Miniati}, F., \& {Viti}, S. 2011, \mnras,
  414, 1705

\bibitem[{{Papadopoulos} {et~al.}(2014){Papadopoulos}, {Zhang}, {Xilouris},
  {et~al.}}]{papa14}
{Papadopoulos}, P.~P., {Zhang}, Z.-Y., {Xilouris}, E.~M., {et~al.} 2014, ApJ,
  788, 153

\bibitem[{{Pentericci} {et~al.}(2016){Pentericci}, {Carniani}, {Castellano},
  {Fontana}, {Maiolino}, {Guaita}, {Vanzella}, {Grazian}, {Santini}, {Yan},
  {Cristiani}, {Conselice}, {Giavalisco}, {Hathi}, \&
  {Koekemoer}}]{pentericci16}
{Pentericci}, L., {Carniani}, S., {Castellano}, M., {et~al.} 2016, \apjl, 829,
  L11

\bibitem[{{Pineda} {et~al.}(2014){Pineda}, {Langer}, \& {Goldsmith}}]{pineda14}
{Pineda}, J.~L., {Langer}, W.~D., \& {Goldsmith}, P.~F. 2014, \aap, 570, A121

\bibitem[{{Pirzkal} {et~al.}(2007){Pirzkal}, {Malhotra}, {Rhoads}, \&
  {Xu}}]{pirzkal07}
{Pirzkal}, N., {Malhotra}, S., {Rhoads}, J.~E., \& {Xu}, C. 2007, \apj, 667, 49

\bibitem[{{Planck Collaboration} {et~al.}(2016){Planck Collaboration}, {Ade},
  {Aghanim}, {Arnaud}, {Ashdown}, {Aumont}, {Baccigalupi}, {Banday},
  {Barreiro}, {Bartlett}, \& et~al.}]{planck16}
{Planck Collaboration}, {Ade}, P.~A.~R., {Aghanim}, N., {et~al.} 2016, \aap,
  594, A13

\bibitem[{{Popping} {et~al.}(2016){Popping}, {van Kampen}, {Decarli}, {Spaans},
  {Somerville}, \& {Trager}}]{popping16}
{Popping}, G., {van Kampen}, E., {Decarli}, R., {et~al.} 2016, \mnras, 461, 93

\bibitem[{{Rahmati} {et~al.}(2013){Rahmati}, {Pawlik}, {Rai{\v
  c}evi$\grave{c}$}, \& {Schaye}}]{rahmati13}
{Rahmati}, A., {Pawlik}, A.~H., {Rai{\v c}evi$\grave{c}$}, M., \& {Schaye}, J.
  2013, \mnras, 430, 2427

\bibitem[{{Rigopoulou} {et~al.}(2014){Rigopoulou}, {Hopwood}, {Magdis},
  {Thatte}, {Swinyard}, {Farrah}, {Huang}, {Alonso-Herrero}, {Bock},
  {Clements}, {Cooray}, {Griffin}, {Oliver}, {Pearson}, {Riechers}, {Scott},
  {Smith}, {Vaccari}, {Valtchanov}, \& {Wang}}]{rigopoulou14}
{Rigopoulou}, D., {Hopwood}, R., {Magdis}, G.~E., {et~al.} 2014, \apjl, 781,
  L15

\bibitem[{{Robitaille}(2011)}]{robitaille11a}
{Robitaille}, T.~P. 2011, \aap, 536, A79

\bibitem[{{Robitaille} {et~al.}(2012){Robitaille}, {Churchwell}, {Benjamin},
  {Whitney}, {Wood}, {Babler}, \& {Meade}}]{robitaille12a}
{Robitaille}, T.~P., {Churchwell}, E., {Benjamin}, R.~A., {et~al.} 2012, \aap,
  545, A39

\bibitem[{{R{\"o}llig} {et~al.}(2006){R{\"o}llig}, {Ossenkopf}, {Jeyakumar},
  {Stutzki}, \& {Sternberg}}]{rollig06}
{R{\"o}llig}, M., {Ossenkopf}, V., {Jeyakumar}, S., {Stutzki}, J., \&
  {Sternberg}, A. 2006, \aap, 451, 917

\bibitem[{{Rosenberg} {et~al.}(2015){Rosenberg}, {van der Werf}, {Aalto},
  {Armus}, {Charmandaris}, {D{\'{\i}}az-Santos}, {Evans}, {Fischer}, {Gao},
  {Gonz{\'a}lez-Alfonso}, {Greve}, {Harris}, {Henkel}, {Israel}, {Isaak},
  {Kramer}, {Meijerink}, {Naylor}, {Sanders}, {Smith}, {Spaans}, {Spinoglio},
  {Stacey}, {Veenendaal}, {Veilleux}, {Walter}, {Wei{\ss}}, {Wiedner}, {van der
  Wiel}, \& {Xilouris}}]{rosenberg15}
{Rosenberg}, M.~J.~F., {van der Werf}, P.~P., {Aalto}, S., {et~al.} 2015, \apj,
  801, 72

\bibitem[{{Rosolowsky}(2005)}]{rosolowsky05}
{Rosolowsky}, E. 2005, \pasp, 117, 1403

\bibitem[{{Sanders} {et~al.}(2015){Sanders}, {Shapley}, {Kriek}, {Reddy},
  {Freeman}, {Coil}, {Siana}, {Mobasher}, {Shivaei}, {Price}, \& {de
  Groot}}]{sanders15}
{Sanders}, R.~L., {Shapley}, A.~E., {Kriek}, M., {et~al.} 2015, \apj, 799, 138

\bibitem[{{Sargsyan} {et~al.}(2014){Sargsyan}, {Samsonyan}, {Lebouteiller},
  {Weedman}, {Barry}, {Bernard-Salas}, {Houck}, \& {Spoon}}]{sargsyan14}
{Sargsyan}, L., {Samsonyan}, A., {Lebouteiller}, V., {et~al.} 2014, \apj, 790,
  15

\bibitem[{{Schaerer} {et~al.}(2015){Schaerer}, {Boone}, {Zamojski}, {Staguhn},
  {Dessauges-Zavadsky}, {Finkelstein}, \& {Combes}}]{schaerer15}
{Schaerer}, D., {Boone}, F., {Zamojski}, M., {et~al.} 2015, \aap, 574, A19

\bibitem[{{Schmidt}(1959)}]{schmidt59}
{Schmidt}, M. 1959, \apj, 129, 243

\bibitem[{{Scoville} {et~al.}(2016){Scoville}, {Sheth}, {Aussel}, {Vanden
  Bout}, {Capak}, {Bongiorno}, {Casey}, {Murchikova}, {Koda},
  {{\'A}lvarez-M{\'a}rquez}, {Lee}, {Laigle}, {McCracken}, {Ilbert}, {Pope},
  {Sanders}, {Chu}, {Toft}, {Ivison}, \& {Manohar}}]{scoville16}
{Scoville}, N., {Sheth}, K., {Aussel}, H., {et~al.} 2016, \apj, 820, 83

\bibitem[{{Seon} {et~al.}(2011){Seon}, {Edelstein}, {Korpela},
  {et~al.}}]{seon11}
{Seon}, K.-I., {Edelstein}, J., {Korpela}, E., {et~al.} 2011, \apjs, 196, 15

\bibitem[{{Smit} {et~al.}(2017){Smit}, {Bouwens}, {Carniani}, {Oesch},
  {Labb{\'e}}, {Illingworth}, {van der Werf}, {Bradley}, {Gonzalez}, {Hodge},
  {Holwerda}, \& {Maiolino}}]{smit17}
{Smit}, R., {Bouwens}, R.~J., {Carniani}, S., {et~al.} 2017, ArXiv e-prints,
  arXiv:1706.04614

\bibitem[{{Smith} {et~al.}(2017){Smith}, {Croxall}, {Draine}, {De Looze},
  {Sandstrom}, {Armus}, {Beir{\~a}o}, {Bolatto}, {Boquien}, {Brandl},
  {Crocker}, {Dale}, {Galametz}, {Groves}, {Helou}, {Herrera-Camus}, {Hunt},
  {Kennicutt}, {Walter}, \& {Wolfire}}]{smith17}
{Smith}, J.~D.~T., {Croxall}, K., {Draine}, B., {et~al.} 2017, \apj, 834, 5

\bibitem[{{Somerville} \& {Dav{\'e}}(2015)}]{somerville15b}
{Somerville}, R.~S., \& {Dav{\'e}}, R. 2015, \araa, 53, 51

\bibitem[{{Speagle} {et~al.}(2014){Speagle}, {Steinhardt}, {Capak}, \&
  {Silverman}}]{speagle14}
{Speagle}, J.~S., {Steinhardt}, C.~L., {Capak}, P.~L., \& {Silverman}, J.~D.
  2014, \apjs, 214, 15

\bibitem[{{Springel}(2005)}]{springel05}
{Springel}, V. 2005, \mnras, 364, 1105

\bibitem[{{Stacey} {et~al.}(1991){Stacey}, {Geis}, {Genzel}, {Lugten},
  {Poglitsch}, {Sternberg}, \& {Townes}}]{stacey91}
{Stacey}, G.~J., {Geis}, N., {Genzel}, R., {et~al.} 1991, \apj, 373, 423

\bibitem[{{Stacey} {et~al.}(2010){Stacey}, {Hailey-Dunsheath}, {Ferkinhoff},
  {Nikola}, {Parshley}, {Benford}, {Staguhn}, \& {Fiolet}}]{stacey10}
{Stacey}, G.~J., {Hailey-Dunsheath}, S., {Ferkinhoff}, C., {et~al.} 2010, \apj,
  724, 957

\bibitem[{{Sturm} {et~al.}(2010){Sturm}, {Verma}, {Graci{\'a}-Carpio},
  {Hailey-Dunsheath}, {Contursi}, {Fischer}, {Gonz{\'a}lez-Alfonso},
  {Poglitsch}, {Sternberg}, {Genzel}, {Lutz}, {Tacconi}, {Christopher}, \& {de
  Jong}}]{sturm10}
{Sturm}, E., {Verma}, A., {Graci{\'a}-Carpio}, J., {et~al.} 2010, \aap, 518,
  L36

\bibitem[{{Turk} {et~al.}(2011){Turk}, {Smith}, {Oishi}, {Skory}, {Skillman},
  {Abel}, \& {Norman}}]{turk11a}
{Turk}, M.~J., {Smith}, B.~D., {Oishi}, J.~S., {et~al.} 2011, \apjs, 192, 9

\bibitem[{{Vallini} {et~al.}(2013){Vallini}, {Gallerani}, {Ferrara}, \&
  {Baek}}]{vallini13}
{Vallini}, L., {Gallerani}, S., {Ferrara}, A., \& {Baek}, S. 2013, \mnras, 433,
  1567

\bibitem[{{Vallini} {et~al.}(2015){Vallini}, {Gallerani}, {Ferrara},
  {Pallottini}, \& {Yue}}]{vallini15}
{Vallini}, L., {Gallerani}, S., {Ferrara}, A., {Pallottini}, A., \& {Yue}, B.
  2015, \apj, 813, 36

\bibitem[{{Venemans} {et~al.}(2012){Venemans}, {McMahon}, {Walter},
  {et~al.}}]{venemans12}
{Venemans}, B.~P., {McMahon}, R.~G., {Walter}, F., {et~al.} 2012, \apjl, 751,
  L25

\bibitem[{{Vladilo}(2004)}]{vladilo04}
{Vladilo}, G. 2004, \aap, 421, 479

\bibitem[{{Wang} {et~al.}(2013{\natexlab{a}}){Wang}, {Wagg}, {Carilli},
  {et~al.}}]{wang13}
{Wang}, R., {Wagg}, J., {Carilli}, C.~L., {et~al.} 2013{\natexlab{a}}, \apj,
  773, 44

\bibitem[{{Wang} {et~al.}(2013{\natexlab{b}}){Wang}, {Wagg}, {Carilli},
  {Walter}, {Lentati}, {Fan}, {Riechers}, {Bertoldi}, {Narayanan}, {Strauss},
  {Cox}, {Omont}, {Menten}, {Knudsen}, {Neri}, \& {Jiang}}]{wang13a}
---. 2013{\natexlab{b}}, \apj, 773, 44

\bibitem[{{Webber}(1998)}]{webber98}
{Webber}, W.~R. 1998, \apj, 506, 329

\bibitem[{{Willott} {et~al.}(2015){Willott}, {Carilli}, {Wagg}, \&
  {Wang}}]{willott15}
{Willott}, C.~J., {Carilli}, C.~L., {Wagg}, J., \& {Wang}, R. 2015, \apj, 807,
  180

\bibitem[{{Willott} {et~al.}(2013){Willott}, {Omont}, \&
  {Bergeron}}]{willott13}
{Willott}, C.~J., {Omont}, A., \& {Bergeron}, J. 2013, \apj, 770, 13

\bibitem[{{Wiseman} {et~al.}(2017){Wiseman}, {Schady}, {Bolmer}, {Kr{\"u}hler},
  {Yates}, {Greiner}, \& {Fynbo}}]{wiseman17}
{Wiseman}, P., {Schady}, P., {Bolmer}, J., {et~al.} 2017, \aap, 599, A24

\bibitem[{{Yang} {et~al.}(2017){Yang}, {Malhotra}, {Gronke}, {Rhoads},
  {Leitherer}, {Wofford}, {Jiang}, {Dijkstra}, {Tilvi}, \& {Wang}}]{yang17}
{Yang}, H., {Malhotra}, S., {Gronke}, M., {et~al.} 2017, ArXiv e-prints,
  arXiv:1701.01857

\bibitem[{{Zhao} {et~al.}(2016){Zhao}, {Yan}, \& {Tsai}}]{zhao16}
{Zhao}, Y., {Yan}, L., \& {Tsai}, C.-W. 2016, \apj, 824, 146

\end{thebibliography}

\end{document}